\documentclass[reqno]{amsart}

\usepackage{amsthm,amsmath,amsfonts,amssymb,epsfig,graphicx,latexsym,float,color}
\usepackage{pgfplots}
\usepackage{xcolor}
\usetikzlibrary{plotmarks}
\usetikzlibrary{external}
\newtheorem{definition}{Definition}

\newtheorem{remark}[definition]{Remark}
\newtheorem{assumption}[definition]{Assumption}

\newtheorem{system}[definition]{System}

\newtheorem{algorithm}[definition]{Algorithm}

\begin{document}

\title[Numerics for Fiber Spinning with Evaporation Effects in Airflows]{An Efficient Numerical Framework for Fiber Spinning Scenarios with Evaporation Effects in Airflows}

\author[Wieland et al.]{Manuel Wieland$^{1}$}
\author[]{Walter Arne$^{2}$}
\author[]{Robert Fe\ss ler$^{2}$}
\author[]{Nicole Marheineke$^{1}$}
\author[]{Raimund Wegener$^{2}$}

\date{\today\\
$^1$ Universit\"at Trier, Lehrstuhl Modellierung und Numerik, Universit\"atsring 15, D-54296 Trier, Germany\\
$^2$ Fraunhofer ITWM, Fraunhofer Platz 1, D-67663 Kaiserslautern, Germany}

\begin{abstract}
In many spinning processes, as for example in dry spinning, solvent evaporates out of the spun jets and leads to thinning and solidification of the produced fibers. Such production processes are significantly driven by the interaction of the fibers with the surrounding airflow. Faced with industrial applications producing up to several hundred fibers simultaneously, the direct numerical simulation of the three-dimensional multiphase, multiscale problem is computationally extremely demanding and thus in general not possible. In this paper, we hence propose a dimensionally reduced, efficiently evaluable fiber model that enables the realization of fiber-air interactions in a two-way coupling with airflow computations. For viscous dry spinning of an uni-axial two-phase flow, we deduce one-dimensional equations for fiber velocity and stress from cross-sectional averaging and combine them with two-dimensional advection-diffusion equations for polymer mass fraction and temperature revealing the radial effects that are observably present in experiments. For the numerical treatment of the resulting parametric boundary value problem composed of one-dimensional ordinary differential equations and two-dimensional partial differential equations we develop an iterative coupling algorithm. Thereby, the solution of the advection-diffusion equations is implicitly given in terms of Green's functions and leads for the surface values to Volterra integral equations of second kind with singular kernel, which we can solve very efficiently by the product integration method. For the ordinary differential equations a suitable collocation-continuation procedure is presented. Compared with the referential solution of a three-dimensional setting, the numerical results are very convincing. They provide a good approximation while drastically reducing the computational time. This efficiency allows the two-way coupled simulation of industrial dry spinning in airflows for which we present results for the first time in literature. 
\end{abstract}

\maketitle

\noindent
{\sc Keywords.} dry spinning, fiber dynamics, parametric boundary value problem, homotopy method, heat and mass transfer, integral equations\\
{\sc AMS-Classification.} 34B08, 68U20, 35Q79, 76-XX

\setcounter{equation}{0} \setcounter{figure}{0} \setcounter{table}{0}
\section{Introduction}

\begin{figure}[!htb]
\centering
\includegraphics[height=7.5cm]{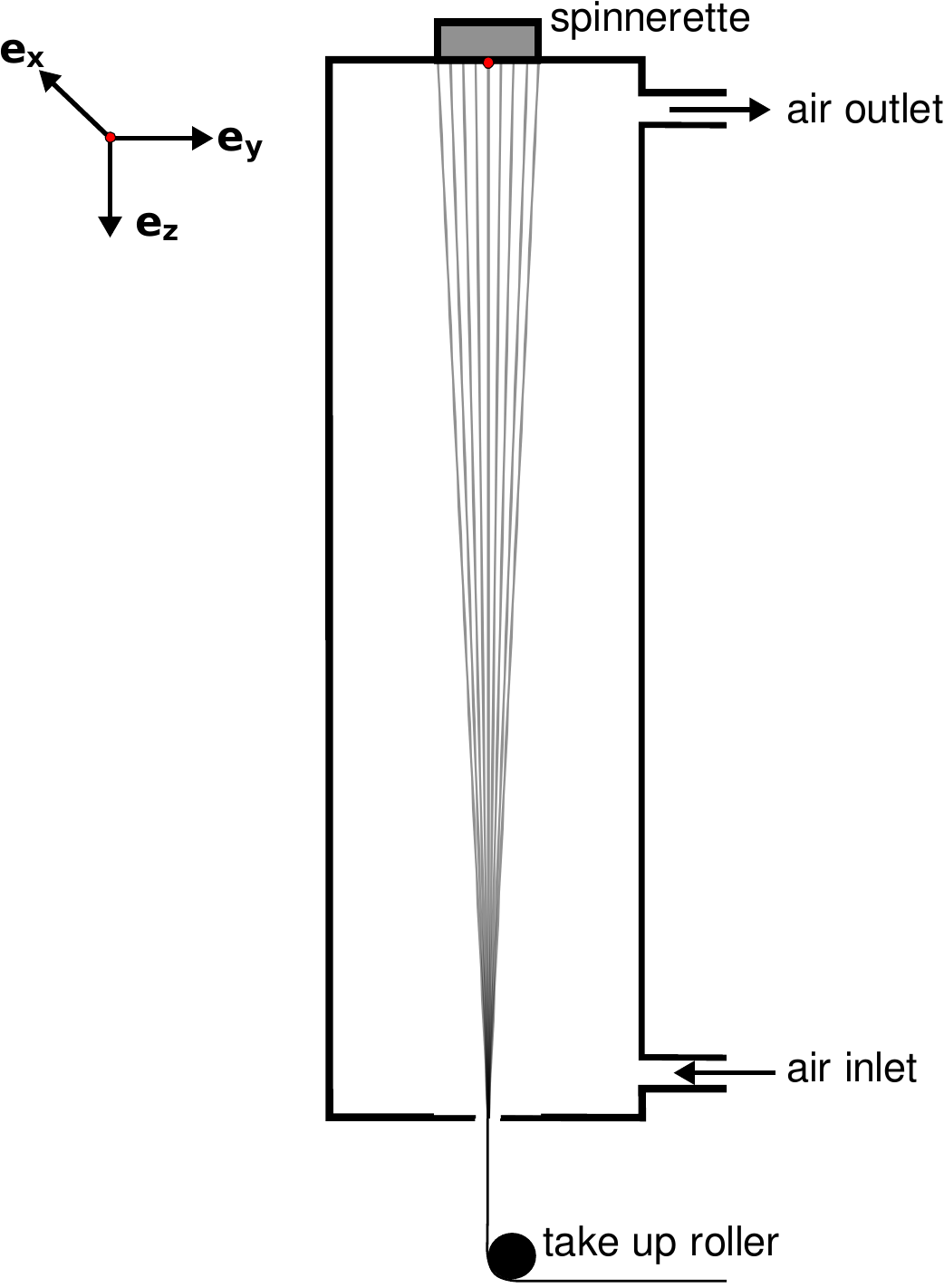}
\caption{Sketch of a dry spinning device with take up of the produced fibers at the bottom side of the spinning column.}\label{sec:intro_fig:apparatus}
\end{figure}

Dry spinning is a widely used production method for fibers, which consist of polymers that must be dissolved in solvent, as for example cellulose acetate, polyacrylonitrile, polyurethane, benzene and many more. As schematized in Fig.~\ref{sec:intro_fig:apparatus} dry spinning devices essentially consist of jet nozzles (spinneret) through which a polymer diluent solution is fed at constant and controllable rate through up to several hundred holes into a spinning column. From these holes of the spinneret the fibers form out, are pulled down by gravity and either laid down on a conveyor belt or drawn down by a take up roller. In the spinning column the fibers are dried by a heated airflow, which usually is blown in at the bottom of the column and exhausts at the top, such that the air forms a counter-current to the fiber flow direction. During this drying process solvent evaporates out of the jets and leads to thinning and solidification of the spun fibers.\par

In the last 50 years, several work has been spent to examine the dry spinning process. One of the first mathematical models was developed by \cite{fok:p:1967}, who employed a species transport equation to predict the solvent concentration in a combined theoretical experimental study. A comprehensive one-dimensional dry spinning model was introduced by \cite{ohzawa:p:1969, ohzawa:p:1970}. There, averaged mass balance equations for the two species (polymer and diluent) as well as averaged momentum and energy balance equations were considered. The balances for solvent concentration and temperature were found to depend crucially on the values at the fiber boundary, which involves the need for radial resolution of the process. This was taken into account through two-dimensional balance equations. A coupling was realized by adjusting corresponding boundary conditions for surface values of the fiber. In \cite{brazinsky:p:1975} theoretical results based on a two-dimensional model, with concentration-dependent thermal conductivity as well as concentration- and temperature-dependent diffusivity and viscosity, were compared with an experimental setup of a cellulose acetate/acetone system. The forming of radial profiles for polymer concentration (mass fraction) and temperature in the fiber was observed. A complete combined theoretical and experimental study of dry spinning of a cellulose acetate/acetone system was presented in \cite{sano:p:2001}, where detailed laws for the concentration- and temperature-dependence of physical properties including diffusivity, elongational viscosity, heat and mass transfer coefficients are specified and incorporated in a coupled one- and two-dimensional model setup. The mentioned studies used exclusively viscous constitutive equations. The works of \cite{gou:p:2003, gou:p:2004} extended the models by incorporating viscoelastic material laws. Apart from that, the effect of solvent evaporation was taken into account in mathematical models for several other spinning processes like melt spinning \cite{kase:p:1965, zerze:p:2015} and electrospinning \cite{yarin:p:2001, wu:p:2011}.

Until now the existing literature solely considers interactions of the simulated fibers with an airflow that is assumed to be constant or measured in experiments. However, in industrial applications we are faced with devices having spinnerets with densely packed holes producing many fibers (possibly several hundred) simultaneously, so that also the fibers visibly affect the surrounding air. This creates the need of a fully two-way coupled simulation of the dry spun fibers immersed in the airflow, at which we aim with this paper. The direct numerical simulation of the three-dimensional multiphase and multiscale problem is computationally extremely demanding and thus in general not possible. Therefore, we propose a dimensionally reduced fiber model for which we design an efficient algorithm such that the simulation of dry spinning processes with two-way coupled fiber-airflow interactions becomes feasible.

Proceeding from a three-dimensional free boundary value problem for a viscous uni-axial radially symmetric two-phase flow we deduce a dimensionally reduced fiber model. From cross-sectional averaging we obtain one-dimensional equations for fiber velocity and stress that we combine consistently with two-dimensional advection-diffusion equations covering the cross-sectional variations of polymer mass fraction and temperature. For the numerical treatment of the one-dimensional ordinary differential equations and the coupling with the two-dimensional partial differential equations we present a continuation algorithm that allows the fast and robust simulation and automatic navigation through a high-dimensional parameter space. The underlying collocation is performed with a Lobatto IIIa formula (implicit fourth order Runge-Kutta scheme), and the resulting nonlinear system is solved with a Newton method. This procedure has been employed successfully in studies on glass wool manufacturing \cite{arne:p:2011} and electrospinning \cite{arne:p:2018}. All previously named works concerning dry spinning employ some types of finite differences for the solution of the involved two-dimensional problems. Here, such methods cannot be utilized since they are time-consuming and the intended further coupling with airflow simulations easily leads to non-practical computation times. Therefore, we develop an efficient algorithmic procedure, which is based on the analytical solution of the two-dimensional advection-diffusion equations with Green's functions \cite{arpaci:b:1991, cole:b:2010, ozisik:b:2013}. The implicitly given solution expressions result in Volterra integral equations of second kind for the surface values of polymer mass fraction and temperature with singular integration kernel, which can be solved quickly by the product integration method. This efficiency allows the realization of two-way coupled fiber and airflow simulations by the help of momentum (drag), heat and mass exchange models according to the principle that action equals reaction \cite{cibis:p:2014, cibis:p:2017}, similar as in the study on glass wool manufacturing \cite{arne:p:2011}. The numerical results show that our reduced dry spinning model is a very good approximation of the underlying three-dimensional model while saving a huge amount of computation time.

The paper is structured as follows. Starting from a three-dimensional free boundary value problem for viscous dry spinning, we derive our dimensionally reduced (one-two-dimensional) fiber model for polymer mass fraction, mixture velocity, stress and temperature in Sec.~\ref{sec:model}. In Sec.~\ref{sec:numerics} we present a continuation-collocation method for the numerical solution of the involved one-dimensional parametric boundary value problem and the coupling with the two-dimensional model equations as well as a product integration method for the two-dimensional profiles of mass fraction and temperature. In addition, the two-way coupling concept for fiber-air interactions is explained. The computational efficiency of our reduced model as well as its good approximation quality with respect to the original three-dimensional problem are demonstrated in Sec.~\ref{sec:performance}. In Sec.~\ref{sec:example} we consider an industrial dry spinning setup of \cite{sano:p:2001}, for which we present simulation results with two-way coupled fiber-air interactions.

\setcounter{equation}{0} \setcounter{figure}{0} \setcounter{table}{0}
\section{Viscous Fiber Dry Spinning Model}\label{sec:model}

In this section we develop our stationary dry spinning model for a viscous uni-axial fiber in a surrounding airflow. We proceed from a three-dimensional free boundary value problem for the two-phase fiber flow given by balances for polymer and diluent mass as well as momentum and energy. Under the assumptions of radial symmetry and slenderness we deduce a dimensionally reduced model that contains the radial effects of polymer mass fraction and fiber (mixture) temperature in combination with the tangential information of fiber velocity and stress.

\subsection{Three-dimensional model}
The dry spinning of a viscous fiber can be described as a stationary three-dimensional free boundary value problem (BVP) for a Newtonian fluid using a mixture model ansatz according to \cite{manninen:p:1996}, which treats the two phases (polymer and solvent) as interpenetrable continua. The conservation principles for mass are formulated for the two single phases, whereas the balances for momentum and energy are stated for the whole mixture. The basic quantities for the model are the polymer mass fraction $c$, the mixture pressure $p$ [Pa], the mixture velocity $\mathbf{v}$ [m/s] and the mixture temperature $T$ [K]. Let $\Omega \subset \mathbb{R}^3$ be the a priori unknown fiber domain whose boundary $\partial\Omega = \Gamma_\mathrm{in} \cup \Gamma_\mathrm{fr}\cup \Gamma_\mathrm{out}$ consists of the fixed inlet at the nozzle $\Gamma_\mathrm{in}$, the free lateral fiber surface $\Gamma_{\mathrm{fr}}$ and the outlet $\Gamma_\mathrm{out}$. The geometry is specified via the kinematic boundary condition on $\Gamma_{\mathrm{fr}}$ with unit outer normal vector $\boldsymbol{\nu}$. The stationary free boundary value problem for the fiber unknowns $c$, $p$, $\mathbf{v}$, $T$ and $\Omega$ reads
\begin{system}[Three-dimensional free BVP]\label{sec:model_eq:3d_model}~\\
Abbreviations:
\begin{equation*}
\rho_p=c\rho, \qquad \rho_d=(1-c)\rho,
\end{equation*}
Balance laws in $\Omega$: 
\begin{equation*}
\begin{aligned} 
 \nabla\cdot(\rho_p\mathbf{v}) &= 0,\\
 \nabla\cdot(\rho_d\mathbf{v}) &= \nabla\cdot\left(\rho D\nabla\left(\frac{\rho_d}{\rho}\right)\right),\\
\nabla\cdot(\rho\mathbf{v}\otimes\mathbf{v})
- \nabla\cdot\left(\rho D\left(\mathbf{v}\otimes\nabla\left(\frac{\rho_d}{\rho}\right) + \nabla\left(\frac{\rho_d}{\rho}\right)\otimes\mathbf{v}\right)\right) &= \nabla\cdot\boldsymbol{\Sigma}^{\mathrm{T}} + \mathbf{g},&\\
\nabla\cdot(\rho h\mathbf{v}) - \nabla\cdot\left(h_d^0 \rho D\nabla\left(\frac{\rho_d}{\rho}\right)\right) &= \nabla\cdot(C\nabla T),
\end{aligned}
\end{equation*}
Kinematic, dynamic, mass and heat flux respective boundary conditions on $\Gamma_{\mathrm{fr}}$:
\begin{equation*}
\begin{aligned}
\mathbf{v}\cdot\boldsymbol{\nu} &= 0,\\
\boldsymbol{\Sigma}\cdot\boldsymbol{\nu} &= \mathbf{f}_\star,\\
-\rho D\nabla\left(\frac{\rho_d}{\rho}\right) \cdot \boldsymbol{\nu} &= j, && j=\gamma\left(\frac{\rho_d}{\rho}-\frac{\rho_{d,\star}}{\varrho}\right),\\
-C\nabla T \cdot \boldsymbol{\nu} &= \alpha(T-T_\star)+  j(\delta-h_d^0),
\end{aligned}
\end{equation*}
Inlet boundary conditions on $\Gamma_{\mathrm{in}}$:
\begin{equation*}
\mathbf{v} = \mathbf{v}_{0},\qquad
\rho_d = \rho_{d,0},\qquad
T = T_{0},
\end{equation*}
Outlet boundary condition on $\Gamma_{\mathrm{out}}$:
\begin{equation*}
\mathbf{v} = \mathbf{v}_{1},
\end{equation*}
Constitutive laws:
\begin{equation*}
\begin{aligned}
\rho^{-1}=c(\rho_p^0)^{-1} +&(1- c)(\rho_d^0)^{-1},\qquad h=c h_p^0+(1-c)h_d^0, 
\\ \boldsymbol{\Sigma} &= -p \mathbf{I} + \mu (\nabla\mathbf{v} + (\nabla\mathbf{v})^{\mathrm{T}}).
\end{aligned}
\end{equation*}
\end{system}

The assumption of an ideal mixture leads to the constitutive laws for the mixture density $\rho$ [kg/m$^3$] and the mixture enthalpy $h$ [J/kg]. Thereby, the material densities $\rho_p^0$ and $\rho_d^0$ as well as the enthalpies $h_p^0$ and $h_d^0$ of pure polymer and diluent, respectively, are temperature-dependent parameters of the model that have to specified. The temperature derivatives of $h_p^0$, $h_d^0$ and $h$ are in particular the specific heat capacities $q_p^0$, $q_d^0$ and $q$ [J/(kgK)] for constant pressure, yielding $q=c q_p^0+(1-c)q_d^0$. The partial densities $\rho_p$ and $\rho_d$ of polymer and diluent in the mixture are defined in terms of mass fraction $c$ and mixture density $\rho$. The mixture velocity $\mathbf{v}$ is the velocity of the polymer phase, and the diffusive velocity of the diluent in the polymer phase is described by a version of Fick's law, which is linear with respect to the diluent mass fraction $\rho_d/\rho = 1-c$, using the diffusion coefficient $D$ [m$^2$/s]. The thermal conductivity is denoted by $C$ [W/(mK)]. For the Newtonian stress tensor $\boldsymbol{\Sigma}$ [Pa] with dynamic mixture viscosity $\mu$ [Pa\,s] we assume incompressibility. The parameters $D$, $C$ and $\mu$ might depend on mass fraction $c$ and temperature $T$. Apart from body forces $\mathbf{g}$ [N/m$^3$] due to gravity, surface forces $\mathbf{f}_\star$ [Pa] due to the surrounding airflow are considered. 

At the lateral fiber surface the diluent density has a jump due to the solvent evaporation, moreover, it changes rapidly in the boundary layer that the surrounding air forms around the fiber. The diluent mass flux $j$ [kg/(m$^2$s)] in the aerodynamic boundary layer can be modeled by the difference of the diluent density in the air at the fiber surface $\varsigma$ [kg/m$^3$] and away from the fiber $\rho_{d,\star}$ [kg/m$^3$] with convective mass transfer coefficient $\beta$ [m/s], i.e., $j=\beta(\varsigma-\rho_{d,\star})$. In System~\ref{sec:model_eq:3d_model} we particularly use a formulation for $j$ in terms of the mass fraction associated transfer coefficient $\gamma=\beta \varrho$ [kg/(m$^2$s)] with $\varrho=\varsigma \rho/\rho_d$ [kg/m$^3$]. The temperature is continuous at the fiber surface, whereas the heat flux has a jump because of the heat exchange in the air due to the solvent evaporation with evaporation enthalpy $\delta$ [J/kg] of the diluent. In the aerodynamic boundary layer the heat flux is described -- analogously to the mass flux -- by the difference of the temperature at the fiber surface and away from the fiber $T_\star$ [K] with heat transfer coefficient $\alpha$ [W/(m$^2$K)]. The parameters $\delta$ and $\varsigma$  might be functions of $c$ and $T$, whereas the transfer coefficients $\alpha$ and $\beta$ depend on the state of the surrounding airflow and especially on the relative velocity between fiber and airflow. In System~\ref{sec:model_eq:3d_model} the surrounding airflow is assumed to be known in the sense that the quantities $\mathbf{f}_\star$, $\rho_{d,\star}$ and $T_\star$ as well as the airflow dependencies of $\alpha$ and $\beta$ are given for each point of the fiber surface.

\begin{remark}[Mass transfer]
The used formulation of the diluent mass transfer $j$ is motivated from the fact that at the fiber surface the diluent density in the air is mainly linearly proportional to the diluent density in the fiber, i.e., $\varsigma \sim \rho_d$.  Setting $\varsigma=(\rho_d / \rho) \varrho$ separates the linear part from the remainder $\varrho$. The mass transfer is principally driven from the linear part in terms of a Robin-type boundary condition, whereas the remainder $\varrho$ is incorporated in the transfer coefficient $\gamma=\beta \varrho$. This splitting might allow a different treatment of the terms, which becomes essentially for our numerics.
\end{remark}

\begin{remark}[Incompressibility]
Considering the mass balance equations for polymer and diluent, assuming constant material densities $\rho_p^0$, $\rho_d^0$  and using the density relation for an ideal mixture yields
\begin{equation*}
\nabla\cdot\left(\mathbf{v} - \frac{\rho}{\rho_d^0} D\nabla\left(\frac{\rho_d}{\rho}\right)\right) = 0,
\end{equation*}
i.e., the flow field is not exactly incompressible with respect to the mixture velocity field $\mathbf{v}$. Nevertheless, we assume an incompressible form of the Newtonian stress tensor $\boldsymbol{\Sigma}$, which is justified by very small diffusive velocities being present in industrial applications.

Alternatively, the mixture model theory might also allow the definition of the mixture velocity $\mathbf{v}$ as linear combination of each singular phase velocity in such a way that $\nabla\cdot\mathbf{v} = 0$ holds. But that ansatz yields diffusive terms in the polymer mass balance equation. Since we aim for a constant polymer flux over the fiber cross-sections, we do not employ it.
\end{remark}

\subsection{Dimensionally reduced model} 
In slender body theory a viscous fiber is usually described by a one-dimensional model of the cross-sectionally averaged balances that are closed by either heuristic assumptions on the radial profiles or asymptotic analysis (see e.g., \cite{panda:p:2008, marheineke:p:2009} and references within). However, for dry spinning the radial resolution of mass fraction and temperature is essential due to the evaporation \cite{ohzawa:p:1969, ohzawa:p:1970}. Combining the relevant two-dimensional aspects with one-dimensional averaged balances, we propose a dimensionally reduced model that is of good approximation quality and very efficiently evaluable.

Considering a single slender fiber we introduce a fixed orthonormal basis $\{\mathbf{e_x},\mathbf{e_y},\mathbf{e_z}\} \subset \mathbb{R}^3$. The body force satisfies $\mathbf{g} = \rho g\mathbf{e_z}$ with gravitational acceleration $g$ and the fiber midline is characterized by a constant unit tangent $\boldsymbol{\tau}$ with $\boldsymbol{\tau} = \mathbf{e_z}$. In addition, we impose the following assumptions on setting and relevant effects.
\begin{assumption}[Setting and fiber geometry]\label{sec:model_assumptions1}
The setting is assumed to be uni-axial, radially symmetric for all fiber quantities and model parameters. In particular, the fiber domain $\Omega$ is given by the fiber length $L$ and the smooth radius function $R:[0,L] \rightarrow \mathbb{R}^+$ such that
\begin{equation*}
\Omega = \{\mathbf{x}=\{x,y,z\}\in\mathbb{R}^3~|~(x,y)\in\mathcal{A}(z),~z\in[0,L]\},
\end{equation*}
with cross-sections
\begin{equation*}
\mathcal{A}(z) = \{(x,y)\in\mathbb{R}^2\big|~ x = r\cos\varphi,~y = r\sin\varphi,~r\in[0,R(z)],~\varphi\in[0,2\pi)\}.
\end{equation*}
\end{assumption}
Then, the lateral free fiber surface $\Gamma_{\mathrm{fr}}$ can be parameterized in terms of the bijective function $\xi:[0,2\pi)\times[0,L] \rightarrow \Gamma_{\mathrm{fr}} \subset\mathbb{R}^3$ with $\xi(\varphi,z) = (R(z)\cos\varphi,R(z)\sin\varphi,z)$ and its outer normal vector $\boldsymbol{\nu}$ is  given by 
\begin{equation}\label{sec:model_eq:nu}
\boldsymbol{\nu}(\varphi,z) = \frac{\partial_\varphi \xi \times \partial_z \xi}{||\partial_\varphi \xi \times \partial_z \xi||}(\varphi,z) = \frac{1}{\sqrt{1 + (\partial_z R(z))^2}} (\cos\varphi,\sin\varphi, -\partial_z R(z)).
\end{equation}
Obviously, $\Gamma_{\mathrm{in}} = \mathcal{A}(0)$ and $\Gamma_{\mathrm{out}} =\mathcal{A}(L)$.
Moreover, it is often convenient to consider the two-dimensional domain $\Omega_{cut} = \{(r,z)\in \mathbb{R}^2 \,|\, r \in (0,R(z)), z\in (0,L)\}\subset \Omega$ with cylindrical coordinates and the orthonormal basis $\{\mathbf{e_r},\mathbf{e_z}\}$.

\begin{assumption}[Diffusive effects]\label{sec:model_assumptions2} The diffusive effects in mass and heat transfer are assumed to be small. Because of the fiber slenderness the diffusive terms that contain radial derivatives are considered of relevance, whereas all others are neglected.
\end{assumption}

\subsubsection{Dimensional reduction}
We proceed from System~\ref{sec:model_eq:3d_model}. For the cross-sec\-tio\-nal averaging we introduce the following notation for any differentiable and integrable scalar-, vector- or tensor-valued function $f$ on $\Omega$ with $\boldsymbol{\nu} = (\nu_x,\nu_y,\nu_z)$ being the unit outer normal vector of $\Gamma_{\mathrm{fr}}$,
\begin{equation*}
\langle f \rangle_{\mathcal{A}(z)} = \int\limits_{\mathcal{A}(z)} f(x,y,z) dxdy,\qquad
\langle f \rangle_{\partial\mathcal{A}(z)} = \int\limits_{\partial\mathcal{A}(z)} \frac{f}{\sqrt{\nu_x^2 + \nu_y^2}} dl.
\end{equation*}
In particular, $|\mathcal{A}(z)|$ and $|\partial \mathcal{A}(z)|$ denote the measures of cross-sectional area and boundary. According to the divergence theorem the averaging rule 
\begin{align*}
\langle \nabla \cdot f \rangle_{\mathcal{A}(z)} = \partial_z \langle f^\mathrm{T}\cdot \mathbf{e_z} \rangle_{\mathcal{A}(z)} + \langle f^\mathrm{T} \cdot \boldsymbol{\nu} \rangle_{\partial\mathcal{A}(z)}
\end{align*}
holds \cite{panda:p:2008}. We apply it on the three-dimensional balances for polymer mass and  momentum, incorporate the boundary conditions on $\Gamma_{\mathrm{fr}}$ and make use of Assumptions~\ref{sec:model_assumptions1} and \ref{sec:model_assumptions2}. In particular, because of the uni-axial, radially symmetric setting, the averaged momentum balance becomes a scalar-valued equation for the $z$-component and the boundary integral simplifies to $\langle f \rangle_{\partial \mathcal{A}}=f |_{\partial \mathcal{A}} |\partial \mathcal{A}|$,
\begin{align*}
\partial_z \langle c\rho v_z \rangle_{\mathcal{A}} = 0,\qquad 
\partial_z \langle \rho v_z^2 \rangle_\mathcal{A} &= 
\partial_z \sigma  + \langle  \rho g\rangle_\mathcal{A} + f_{air} - (j v_z)\big|_{\partial \mathcal{A}} |\partial \mathcal{A}|,
\end{align*}
with $\sigma = \langle (\boldsymbol{\Sigma}\cdot\mathbf{e_z})\cdot\mathbf{e_z}\rangle_{\mathcal{A}}=\langle -p+2\mu \partial_z v_z \rangle_{\mathcal{A}}$ and $f_{air} = \langle \mathbf{f_\star}\cdot\mathbf{e_z} \rangle_{\partial\mathcal{A}}$. Obviously, by the averaging the radial information of the velocity components $v_z = \mathbf{v}\cdot\mathbf{e_z}$, $v_r = \mathbf{v}\cdot\mathbf{e_r}$ and the pressure $p$ gets lost. Therefore, we impose the following assumption that -- as well as Assumption~\ref{sec:model_assumptions2} -- might be justified by asymptotic analysis (see e.g.\ \cite{panda:p:2008} for profiles for $v_z$ and $p$).
\begin{assumption}[Velocity and pressure profiles]\label{sec:model_ansatz:profiles}
The velocity components are considered to have the form $v_z(r,z) = u(z)$ and $v_r(r,z) = r/R(z)v_r|_{\partial\mathcal{A}(z)}(z)$, while the pressure is $p(r,z) = -\mu(c,T)\partial_z u(z)$.
\end{assumption}

All together the one-dimensional averaged polymer mass and momentum balances become
\begin{equation}\label{sec:model_eq:1d}
\begin{aligned}
\partial_z (\langle c\rho(c,T)\rangle_\mathcal{A}\,u) &= 0,\\
\partial_z (\langle\rho(c,T)\rangle_\mathcal{A}\,u^2) &=  \partial_z \sigma + g\langle\rho(c,T) \rangle_{\mathcal A} + f_{air} - u\, j(c,T)\big|_{\partial\mathcal{A}}|\partial \mathcal A|, \qquad \sigma = 3 \langle \mu(c,T)\rangle_{\mathcal{A}} \,\partial_z u,
\end{aligned}
\end{equation}
where $\rho$, $\mu$ and $j$ are functions of mass fraction $c$ and temperature $T$. We conclude the required information about the profiles of $c$ and $T$ from two-dimensional advection-diffusion equations. For this purpose we rewrite the three-dimensional balances for diluent mass and energy of System~\ref{sec:model_eq:3d_model} by employing the total mass balance, the product rule as well as the definition of the specific heat capacities $q$. The resulting equations
\begin{align*}
\mathbf{v}\cdot\nabla c - \frac{c}{\rho}\nabla\cdot(\rho D\nabla c) &= 0,\qquad
\rho q \mathbf{v}\cdot\nabla T + \rho D \nabla h_d^0 \cdot\nabla c - \nabla\cdot(C \nabla T) = 0.
\end{align*}
are then formulated in cylindrical coordinates  on $\Omega_{cut}$, regarding Assumptions~\ref{sec:model_assumptions1}, \ref{sec:model_assumptions2} and \ref{sec:model_ansatz:profiles}. For the radial velocity component we particularly apply the expression 
\begin{align*}
v_r(r,z) = r\frac{\partial_zR(z)}{R(z)}u(z),
\end{align*}
which is concluded by means of the homogeneous kinematic boundary condition on the free lateral fiber surface, $ v_r\big|_{\partial\mathcal{A}} - u\partial_z R = 0$ (cf.~\eqref{sec:model_eq:nu}). So,
\begin{equation}\label{sec:model_eq:2d}
\begin{aligned}
u\,\partial_z c + r\frac{\partial_zR}{R}u\,\partial_rc - \frac{c}{\rho(c,T)}\frac{1}{r}\partial_r(r\rho(c,T)D(c,T)\,\partial_r c) &= 0,\\
\rho(c,T)q(c,T)u\left(\partial_z T + r\frac{\partial_zR}{R}\,\partial_ r T\right) + \rho(c,T) D(c,T)\partial_r h_d^0(T)\, \partial_r c - \frac{1}{r}\partial_r(r\, C(c,T)\,\partial_r T) &= 0.
\end{aligned}
\end{equation}

\subsubsection{Dimensionless formulation}
For the numerics it is convenient to deal with dimensionless model equations. For simplicity we consider here constant inflow profiles, $c(r,0) = c_0 = const$ and $T(r,0) = T_0 = const$. Then, we non-dimensionalize the averaged balances (\ref{sec:model_eq:1d}) and the radial advection-diffusion equations (\ref{sec:model_eq:2d}) using fiber speed $u_0$, temperature $T_0$, density $\rho_0=\rho(c_0,T_0)$, heat capacity $q_0=q(c_0,T_0)$, dynamic viscosity $\mu_0=\mu(c_0,T_0)$, diffusivity $D_0 = D(c_0,T_0)$ and thermal conductivity $C_0 = C(c_0,T_0)$ at the nozzle (inlet) as well as nozzle radius $R_0$ and fiber length $L$ (being equivalent to the device height for a straight uni-axial fiber). We introduce the dimensionless quantities as $\tilde{\mathsf{y}}(\tilde{r},\tilde{z})=\mathsf{y}(\hat{r}\tilde{r},\hat{z}\tilde{z})/\hat y$ where the reference values are marked with $\hat{~}$. Besides the already named reference values we choose $\hat{\sigma}=\pi\mu_0 u_0 R_0^2/L$, $\hat{h}=q_0 T_0$, $\hat{\alpha}=\rho_0 u_0 q_0$, $\hat{\gamma}=\rho_0 u_0$ and $\hat{f}_{air} = \mu^2_\star/(2R\rho_\star )$ where we especially use the density $\rho_\star$ [kg/m$^3$] and the dynamic viscosity $\mu_\star$ [Pas] of the air. To keep the notation simple we suppress the label $\tilde{~}$ in the following and write 
$\langle f \rangle_{R^2(z)} = 2\pi\int_0^1 f(r,z)rdr$
for the integrated dimensionless quantity $f$ under the assumption of radial symmetry (Assumption~\ref{sec:model_assumptions1}). The dimensionless one-two-dimensional fiber model is described by System~\ref{sec:model_eq:coupled_sys}.
\begin{system}[One-two-dimensional BVP]\label{sec:model_eq:coupled_sys}~\\
Averaged balance laws, $z\in (0,1)$:
\begin{equation}\label{sec:model_eq:modified_1d}
\begin{aligned}
\partial_z(\langle c\rho(c,T) \rangle_{R^2} u) &= 0,\\
\partial_z(\langle\rho(c,T)\rangle_{R^2}u^2) &= \frac{1}{\mathrm{Re}}\partial_z\sigma + \frac{1}{\mathrm{Fr}^2}\langle \rho(c,T)\rangle_{R^2} + \mathrm{M}\frac{1}{R} f_{air} - \frac{2}{\epsilon}Ru\,j(c,T)\big|_{r=1},
\end{aligned}
\end{equation}
with boundary conditions at inlet $z=0$ and outlet $z=1$:
\begin{align*}
u(0) = 1, \qquad  u(1) = \mathrm{Dr},
\end{align*}
Radial equations, $(r,z)\in (0,1)^2$:
\begin{equation}\label{sec:model_eq:modified_2d}
\begin{aligned}
u\,\partial_z c - \frac{1}{\epsilon\mathrm{Pe_D}}\frac{c}{\rho(c,T)R^2r}\partial_{r}(r\rho(c,T)D(c,T)\,\partial_r c) &= 0,\\
\rho(c,T) q(c,T) u\,\partial_z T + \frac{1}{\epsilon\mathrm{Pe_D}}\frac{\rho(c,T)D(c,T)}{R^2}\partial_r h_d^0(T)\,\partial_r c - \frac{1}{\epsilon\mathrm{Pe_C}}\frac{1}{R^2r}\partial_r(rC(c,T) \,\partial_r T) &= 0,
\end{aligned}
\end{equation}
with boundary conditions at inlet $z=0$, fiber surface $r=1$ and symmetry boundary $r=0$:
\begin{align*}
c\big|_{z=0} &= c_0, \qquad \quad \partial_r c\big|_{r=0} = 0,\\
\frac{1}{\mathrm{Pe_D}}\frac{\big(\rho(c,T)D(c,T)\big)\big|_{r=1}}{R}\partial_r c\big|_{r=1} &= j(c,T)\big|_{r=1}, \qquad \qquad \, \qquad j(c,T) = -\gamma(c,T)(c - c_{ref}(c,T)),\\
T\big|_{z=0} &= 1, \qquad \quad \,\partial_r T\big|_{r=0} = 0,\\
-\frac{1}{\mathrm{Pe_C}}\frac{C(c,T)\big|_{r=1}}{R}\partial_{r} T\big|_{r=1} &= \big(\alpha(T-T_\star) + j(c,T)(\delta(T)-h_d^0(T))\big)\big|_{r=1},
\end{align*}
Constitutive laws:
\begin{equation*}
\begin{aligned}
\rho^{-1}(c,T) &= c\,(\rho_p^0)^{-1}(T) + (1-c)\,(\rho_d^0)^{-1}(T),\qquad &q(c,T) &= cq_p^0(T) + (1-c)q_d^0(T),\\
\sigma &= 3\langle \mu(c,T) \rangle_{R^2}\partial_z u,\qquad & q_d^0(T)&=\partial_T h_d^0(T) .\\
\end{aligned}
\end{equation*}
\end{system}
Thereby, introducing the referential polymer mass fraction $c_{ref} = 1-\rho_{d,\star}/\varrho$ the diluent mass flux $j$ is formulated with respect to the polymer mass fraction. Note that due to the transformation of the radial equations on the unit square the terms containing the radial velocity cancel out. System~\ref{sec:model_eq:coupled_sys} is characterized by seven dimensionless parameters: length ratio $\epsilon$, draw ratio $\mathrm{Dr}$, Peclet numbers $\mathrm{Pe_D}$, $\mathrm{Pe_C}$ as ratio of advection rate to diffusion rate for mass and heat transfer respectively, Reynolds number $\mathrm{Re}$ as ratio of inertial to viscous forces, Froude number $\mathrm{Fr}$ as ratio of inertial to gravitational forces as well as air-drag associated parameter $\mathrm{M}$, i.e.,
\begin{align*}
\epsilon &= \frac{R_0}{L}, \qquad \mathrm{Dr}=\frac{u_1}{u_0}, \qquad \mathrm{Pe_D} = \frac{u_0R_0}{D_0}, \qquad \mathrm{Pe_C} = \frac{\rho_0 q_0 u_0R_0}{C_0}\\
\mathrm{Re} &= \frac{\rho_0u_0L}{\mu_0}, \qquad \mathrm{Fr} = \frac{u_0}{\sqrt{gL}}, \qquad \mathrm{M}(z) = \frac{L}{2\pi \rho_0 u_0^2 R_0^3}\frac{\mu_\star^2}{\rho_\star}\bigg|_z.
\end{align*}
In contrast to the other numbers, the air-drag associated parameter $\mathrm{M}$ is a scalar field due to its dependence on the airflow quantities $\mu_\star$, $\rho_\star$ varying in $z$-direction.

\subsubsection{Modifications for efficient evaluation}
Aiming for the simulation of industrial dry spinning with two-way coupled fiber-air interactions the computation of the fiber model itself must be performed very quickly. In spite of the dimensional reduction, the numerical solving of the nonlinear two-dimensional advection-diffusion equations  \eqref{sec:model_eq:modified_2d} is still too time consuming for such setups. Hence, we propose a modified version of System~\ref{sec:model_eq:coupled_sys} that makes an efficient evaluation possible.

We linearize the advection-diffusion equations \eqref{sec:model_eq:modified_2d} around the cross-sectionally averaged mass fraction $\bar{c}$ and temperature $\bar{T}$, i.e.,
\begin{align*}
\bar{c} = \frac{1}{\pi}\langle c \rangle_{R^2},\qquad \bar{T} = \frac{1}{\pi}\langle T \rangle_{R^2}.
\end{align*}
Apart from the leading terms of zeroth order, we take into account the terms of first order with respect to $c$ in the equation for the mass fraction and, analogously, the terms of first order with respect to $T$ in the equation for the temperature,
\begin{equation}\label{sec:model_eq:simplified2d(1)}
\begin{aligned}
u\partial_z c - \frac{1}{\epsilon\mathrm{Pe_D}}\frac{\bar{c}D(\bar{c},\bar{T})}{R^2}\frac{1}{r}\partial_{r}(r\partial_r c) &= 0,\\
\rho(\bar{c},\bar{T}) q(\bar{c},\bar{T}) u\partial_z T - \frac{1}{\epsilon\mathrm{Pe_C}}\frac{C(\bar{c},\bar{T})}{R^2 r}\partial_r(r\partial_r T) &= 0.
\end{aligned}
\end{equation}
The boundary conditions at the fiber surface $r=1$ are respectively
\begin{align*}
\frac{1}{\mathrm{Pe_D}}\frac{\rho(\bar{c},\bar{T})D(\bar{c},\bar{T})}{R}\partial_r c\big|_{r=1} &= j(c,T)\big|_{r=1},\\
-\frac{1}{\mathrm{Pe_C}}\frac{C(\bar{c},\bar{T})}{R}\partial_{r} T\big|_{r=1} &= \big(\alpha(T-T_\star) + j(c,T)(\delta(T)-h_d^0(T))\big)\big|_{r=1}.
\end{align*}
Through the linearized form of the radial equations \eqref{sec:model_eq:simplified2d(1)}
the evaluation and subsequent averaging of the density profiles in the averaged balance laws \eqref{sec:model_eq:modified_1d} becomes insignificant. Therefore, we apply the simplification
that the  cross-sectionally averaged mixture density and partial polymer density can be well approximated by
\begin{align*}
\langle\rho(c,T)\rangle_{R^2} = \rho(\bar{c},\bar{T})R^2,\qquad \langle c\rho(c,T)\rangle_{R^2} = \bar{c}\rho(\bar{c},\bar{T})R^2.
\end{align*}
Then, the averaged balance laws become
\begin{equation*}
\begin{aligned}
\partial_z(R^2\bar{c}\rho(\bar{c},\bar{T})   u) &= 0,\\
\partial_z(R^2\rho(\bar{c},\bar{T}) u^2) &= \frac{1}{\mathrm{Re}}\partial_z\sigma + \frac{1}{\mathrm{Fr}^2}R^2\rho(\bar{c},\bar{T}) + \mathrm{M}\frac{1}{R} f_{air} - \frac{2}{\epsilon}Ru\,j(c,T)\big|_{r=1}.
\end{aligned}
\end{equation*}
Obviously, we do not face a free boundary value problem any more, as the fiber radius $R$ can be concluded from the constant polymer flux $Q=R^2\bar{c}\rho(\bar{c},\bar{T})u$ using the boundary conditions at the inlet, i.e.,
\begin{align*}
R(z) = \sqrt{ \frac{c_0}{(\bar{c}\rho(\bar{c},\bar{T}) u)|_z}}.
\end{align*}

For the later numerical treatment in Sec.~\ref{sec:numerics} we need suitable initial guesses for the averaged quantities $\bar{c}$ and $\bar{T}$. A good strategy is to solve the radial equations \eqref{sec:model_eq:simplified2d(1)} averaged over the cross-sections, yielding
\begin{equation}\label{sec:model_eq:avg_cT}
\begin{aligned}
\partial_z \bar{c} &= \frac{2}{\epsilon} \frac{\bar{c}^2}{c_0} R\, j(c,T)\big|_{r=1},\\
\partial_z \bar{T} &= -\frac{2}{\epsilon} \frac{\bar{c}}{c_0} \frac{R}{q(\bar{c},\bar{T})}\big(\alpha(T - T_\star) +j(c,T)(\delta(T) - h_d^0(T))\big)\big|_{r=1},
\end{aligned}
\end{equation}
with initial data $\bar{c}(0) = c_0$ and $\bar{T}(0) = 1$. 
Using \eqref{sec:model_eq:avg_cT} we conclude our final fiber model for convective speed $u$, stress $\sigma$, mass fraction $c$ and temperature $T$, cf.\ System~\ref{sec:model_eq:coupled_sys_simpl}.
\begin{system}[Simplified one-two-dimensional BVP]\label{sec:model_eq:coupled_sys_simpl}~\\
One-dimensional equations, $z\in (0,1)$:
\begin{equation}\label{sec:model_eq:simplified_1d}
\begin{aligned}
\partial_z u &= \frac{1}{3\langle\mu(c,T)\rangle_{R^2}}\sigma,\\
\partial_z \sigma &= \frac{\mathrm{Re}}{3}\frac{c_0}{\bar{c}} \frac{1}{\langle\mu(c,T) \rangle_{R^2}}\sigma     - \frac{\mathrm{Re}}{\mathrm{Fr}^2} \frac{c_0}{\bar{c}}\frac{1}{u} - \mathrm{Re} \mathrm{M}\frac{1}{R}f_{air},
\end{aligned}
\end{equation}
with boundary conditions at inlet $z=0$ and outlet $z=1$:
\begin{align*}
u(0) = 1, \qquad  u(1) = \mathrm{Dr},
\end{align*}
Two-dimensional equations, $(r,z)\in (0,1)^2$:
\begin{equation}\label{sec:model_eq:simplified_2d}
\begin{aligned}
u\partial_z c - \frac{1}{\epsilon\mathrm{Pe_D}}\frac{\bar{c}D(\bar{c},\bar{T})}{R^2r}\partial_{r}(r\partial_r c) &= 0,\\
\rho(\bar{c},\bar{T}) q(\bar{c},\bar{T}) u\partial_z T - \frac{1}{\epsilon\mathrm{Pe_C}}\frac{C(\bar{c},\bar{T})}{R^2 r}\partial_r(r\partial_r T) &= 0,
\end{aligned}
\end{equation}
with boundary conditions at inlet $z=0$, fiber surface $r=1$ and symmetry boundary $r=0$:
\begin{align*}
c\big|_{z=0} &= c_0, \qquad \qquad \partial_r c\big|_{r=0} = 0,\\
\frac{1}{\mathrm{Pe_D}}\frac{\rho(\bar{c},\bar{T})D(\bar{c},\bar{T})}{R}\partial_r c\big|_{r=1} &= j(c,T)\big|_{r=1},\qquad \qquad \qquad j(c,T) = -\gamma(c,T)(c - c_{ref}(c,T)),\\
T\big|_{z=0} &= 1, \qquad \qquad \partial_r T\big|_{r=0} = 0,\\
-\frac{1}{\mathrm{Pe_C}}\frac{C(\bar{c},\bar{T})}{R}\partial_{r} T\big|_{r=1} &= \big(\alpha(T-T_\star) + j(c,T)(\delta(T)-h_d^0(T))\big)\big|_{r=1},
\end{align*}
Constitutive laws and geometric relation:
\begin{equation*}
\begin{aligned}
\rho^{-1}(c,T) &= c\,(\rho_p^0)^{-1}(T) + (1-c)\,(\rho_d^0)^{-1}(T),\qquad &q(c,T) &= cq_p^0(T) + (1-c)q_d^0(T),\\
R(z) &= \sqrt{\frac{c_0}{(\bar{c}\rho(\bar{c},\bar{T})u)|_z}},\qquad & q_d^0(T)&=\partial_T h_d^0(T) .\\
\end{aligned}
\end{equation*}
Abbreviations:
\begin{align*}
\bar{c} = \frac{1}{\pi}\langle c \rangle_{R^2},\qquad \qquad \bar{T} = \frac{1}{\pi}\langle T \rangle_{R^2}.
\end{align*}
\end{system}

\begin{remark}[Dynamic viscosity]\label{sec:model_remark:dyn_visc}
In System~\ref{sec:model_eq:coupled_sys_simpl} we could approximate the averaged dynamic viscosity in the same way as the densities, i.e., $\langle \mu(c,T) \rangle_{R^2} = \mu(\bar{c},\bar{T})R^2$. However, for reasons of accuracy we do not employ this simplification, as we will discuss in Sec.~\ref{subsec:closing}.
\end{remark}

\begin{remark}[Closure models]\label{sec:model_remark}
The closing of System \ref{sec:model_eq:coupled_sys_simpl} requires appropriate models for drag force $f_{air}$, heat transfer coefficient $\alpha$, mass transfer coefficient $\gamma$, referential mass fraction $c_{ref}$, diffusion coefficient $D$, thermal conductivity $C$, evaporation enthalpy $\delta$, specific heat capacities $q_p^0$, $q_d^0$, dynamic viscosity $\mu$ as well as material densities $\rho_p^0$, $\rho_d^0$.

For the performance and quality analysis of our dimensionally reduced fiber model in Sec.~\ref{sec:performance} we set all these quantities to be constant, whereas for the setup of an industrial dry spinning apparatus in Sec.~\ref{sec:example} we give detailed models taking the fiber materials and rheological effects into account. The momentum (drag), heat and mass exchange models are particularly based on airflow simulations. According to the principle that action equals reaction two-way coupled fiber-airflow interactions are realized.
\end{remark}

\setcounter{equation}{0} \setcounter{figure}{0} \setcounter{table}{0}
\section{Numerical Scheme}\label{sec:numerics}

The derived uni-axial fiber model for dry spinning of a single fiber (System~\ref{sec:model_eq:coupled_sys_simpl}) is a combined system of one- and two-dimensional model equations. For given polymer mass fraction $c$ and mixture temperature $T$, the one-dimensional equations (\ref{sec:model_eq:simplified_1d}) form together with the boundary conditions a parametric boundary value problem of ordinary differential equations with two variables. The numerical challenge lies in solving the problem for arbitrary parameter settings, which requires suitable initial guesses of the respective solutions. This problem is tackled by a continuation-collocation method that makes the efficient and robust simulation and navigation through a high-dimensional parameter space possible. The solution of the one-dimensional equations is then iteratively coupled with the solution of the two-dimensional equations \eqref{sec:model_eq:simplified_2d}. Before the first coupling iteration the necessary fields of averaged mass fraction $\bar{c}$ and temperature $\bar{T}$ are calculated by the averaged advection-diffusion equations \eqref{sec:model_eq:avg_cT} with $(c,T)|_{r=1}=(\bar c, \bar T)$ as initial guess. A solution of the two-dimensional equations (\ref{sec:model_eq:simplified_2d}) itself can implicitly be given in terms of Green's functions and Volterra integral equations of second kind with singular kernel for the values at the fiber boundary. We employ a suitable product integration method based on the same Lobatto IIIa quadrature formula as the collocation scheme.

The basic coupling algorithm for the solving of System~\ref{sec:model_eq:coupled_sys_simpl} has the following form.
\begin{algorithm}[Fiber solution]~\label{sec:numerics_algo1}
\begin{enumerate}
\item[(1)] Solve the averaged equations \eqref{sec:model_eq:simplified_1d} and \eqref{sec:model_eq:avg_cT} with $(c,T)|_{r=1} = (\bar{c},\bar{T})$ and $\langle \mu(c,T)\rangle_{R^2} = \mu(\bar{c},\bar{T})R^2$ to obtain $\mathbf{y}^{(0)} = (u,\sigma,\bar{c},\bar{T})$.
\item[(2)] For $i\geq 1$, solve \eqref{sec:model_eq:simplified_2d} with given $\mathbf{y}^{(i-1)}$ to obtain $(c,T)^{(i)}$ and afterwards compute the new solution of \eqref{sec:model_eq:simplified_1d} to obtain $(u,\sigma)^{(i)}$, which gives in total $\mathbf{y}^{(i)}$, as long as $\lVert \mathbf{y}^{(i-1)} - \mathbf{y}^{(i)} \rVert > tol$ for given tolerance $tol$.
\end{enumerate}
\end{algorithm}
Finally, the fiber-air interaction is performed by an algorithm that weakly couples the fiber calculation and airflow computation via iterative solving.

\subsection{Collocation method}
To solve a boundary value problem of the form
\begin{align}\label{sec:numerics_eq:bvp}
\frac{\mathrm{d}}{\mathrm{d}z}\mathsf{y}=\mathsf{f}(\mathsf{y}), \qquad \mathsf{g}(\mathsf{y}(0),\mathsf{y}(1))=\mathsf{0},
\end{align} 
we use a three-stage Lobatto IIIa formula as collocation scheme \cite{hairer:b:2009}. It is an implicit Runge-Kutta method. The collocation polynomial provides an once continuously differentiable solution that is fourth-order accurate uniformly in $z \in [0,1]$. Mesh selection and error control are based on the residual of the continuous solution \cite{kierzenka:p:2008}. Thus, we have
\begin{equation}\label{sec:numerics_eq:collocation}
\begin{aligned}
&\mathsf{y}_{i+1}-\mathsf{y}_i - \frac{h_{i+1}}{6}(\mathsf{f}(\mathsf{y}_i) + 4\mathsf{f}(\mathsf{y}_{i+1/2}) + \mathsf{f}(\mathsf{y}_{i+1})) =\mathsf{0},\qquad \mathsf{g}(\mathsf{y}_0,\mathsf{y}_N)=\mathsf{0},\\
&\text{with} \quad \mathsf{y}_{i+1/2} = \frac{1}{2}(\mathsf{y}_{i+1}+\mathsf{y}_i) - \frac{h_{i+1}}{8}(\mathsf{f}(\mathsf{y}_{i+1}) - \mathsf{f}(\mathsf{y}_{i})),
\end{aligned}
\end{equation}
with collocation points $0=z_0<z_1<...<z_N=1$, mesh size $h_i=z_i-z_{i-1}$ and the abbreviation $\mathsf{y}_i=\mathsf{y}(z_i)$. The resulting nonlinear system of $N+1$ equations for $(\mathsf{y}_i)_{i=0,...,N}$ is solved using a Newton method with numerically approximated Jacobian. This is a classical approach that is provided in the software MATLAB\footnote{For details see \texttt{http://www.mathworks.com}.} by the routine \texttt{bvp4c.m}. Its applicability depends on the convergence of the Newton method that is crucially determined by the initial guess. We aim for adapting the initial guess iteratively by means of a continuation method, solving a sequence of slightly varying boundary value problems. 

\subsection{Continuation method}
In the continuation method we embed the boundary value problem of interest \eqref{sec:numerics_eq:bvp} into a family of problems by introducing a continuation parameter tuple $\mathsf{p}\in[0,1]^n$, $n\in \mathbb{N}$
\begin{align*}
\frac{\mathrm{d}}{\mathrm{d}z}\mathsf{y}=\mathsf{\hat f}(\mathsf{y};\mathsf{p}),  \qquad & \mathsf{\hat g}(\mathsf{y}(0),\mathsf{y}(1);\mathsf{p})=\mathsf{0}, && \mathsf{p}\in[0,1]^n,\\
\mathsf{\hat f}(\cdot;\mathsf{\underline 1})=\mathsf{f}, \qquad \mathsf{\hat g}(\cdot,\cdot;\mathsf{\underline 1})=\mathsf{g}, \qquad & \mathsf{\hat f}(\cdot;\mathsf{0})=\mathsf{f}_0, \qquad \mathsf{\hat g}(\cdot,\cdot;\mathsf{0})=\mathsf{g}_0.
\end{align*}
Here, $\mathsf{\underline 1}$ denotes the $n$-dimensional tuple of ones. The functions $\mathsf{f}_0$, $\mathsf{g}_0$ are chosen in such a way that for $\mathsf{p}=\mathsf{0}$ an analytical solution is known. Given this starting solution, we seek for a sequence of parameter tuples $\mathsf{0}=\mathsf{p}_0, \mathsf{p}_1, \ldots, \mathsf{p}_m = \mathsf{\underline 1}$ such that the solution to the respective predecessor boundary value problem provides a good initial guess for the successor. The solution associated to $\mathsf{p}= \mathsf{\underline 1}$ finally belongs to the original system. By help of the continuation parameters certain terms in the ordinary differential equations can be first excluded, then included. Also the boundary conditions can be varied. The core of a robust continuation procedure are the choice of a continuation path and the step size control to navigate through a high-dimensional parameter space. They decide about failure or success because there are not always existing solutions and several ways might be possible. Whereas the choice of the continuation path is problem-specific, the step size control is a general question and we follow an approach that was successfully employed in studies on glass wool manufacturing \cite{arne:p:2011} and electrospinning \cite{arne:p:2018}.  

To explain the used procedure, we consider at first a one-dimensional parameter space $p\in [0,1]$. Proceeding from an initial continuation step size $\Delta p_0$, a boundary value problem is always solved twice by using one full step and two half steps. If the full step requires more Newton iterations than both half steps together or $k_1$-times more collocation points than the second half step, the continuation step is reduced by a factor $k_2$, otherwise it is increased by $k_2$ for the further computation. If the Newton method fails, the step size is reduced by a factor $k_3$ and the computation is repeated. Certainly, it can happen that no solutions exist for $p>p_{crit}\geq 0$, thus the algorithm for the adaptive step size control has a stopping criterion that is based on a minimal step size. In particular, we use $k_1 = 0.1$, $k_2=1.5$, $k_3=10$, $\Delta p_0=10^{-1}$, $\Delta p_{min}=10^{-14}$.

In a high-dimensional parameter space $\mathsf{p}\in [0,1]^n$, $n\gg1$ there might be numerous possible but also impossible continuation paths for a specific model problem. We use a heuristic that groups the continuation parameters and applies to each group the one-dimensional continuation procedure successively as described in the following.

For our dry spinning application we apply different continuation strategies for the two steps in Algorithm~\ref{sec:numerics_algo1}. We denote Step (1) as initialization step, whereas Step (2) is called the coupling step.

\subsubsection{Initialization step}
We introduce continuation parameters for the viscous, gravitational and aerodynamic forces as well as for the draw ratio $p_\mathrm{Re}$, $p_\mathrm{Fr}$, $p_\mathrm{M}$, $p_\mathrm{Dr} \in [0,1]$. This means we replace the global Reynolds number $\mathrm{Re}$ by the term $p_\mathrm{Re}\mathrm{Re}+(1-p_\mathrm{Re})\mathrm{Re}_0$, analogously for $\mathrm{Fr}$, $\mathrm{M}$ and $\mathrm{Dr}$ with starting values  $\mathrm{Re}_0 = 10$, $\mathrm{Fr}_0 = 100$, $\mathrm{M}_0 = 10^{-4}$, $\mathrm{Dr}_0 = 1$. Furthermore, we introduce continuation parameters $p_1$, $p_2$ on the right hand side of the equations \eqref{sec:model_eq:avg_cT} to control the effect of heat and mass transfer over the fiber boundary,
\begin{equation*}
\begin{aligned}
\partial_z \bar{c} &= \frac{1}{\epsilon} \frac{\bar{c}^2}{c_0} 2R j(\bar{c},\bar{T})p_1,\\
\partial_z \bar{T} &= -\frac{1}{\epsilon} \frac{\bar{c}}{c_0} \frac{2R}{q(\bar{c},\bar{T})}\big(\alpha(\bar{T} - T_\star) +j(\bar{c},\bar{T})(\delta(\bar{T}) - h_d^0(\bar{T}))\big)p_2.
\end{aligned}
\end{equation*}
Consequently, we embed the one-dimensional fiber model into a family of boundary value problems in a six-dimensional parameter space, $\mathsf{p}=(p_1,p_2, p_\mathrm{Re},p_\mathrm{Fr}, p_\mathrm{M},p_\mathrm{Dr})\in [0,1]^{6}$. The choice of the continuation parameters allows the decrease and increase of mass and heat transfer as well as of viscosity, gravity, air drag and drawing effects that mainly dominate the fiber dynamics.

The starting solution for the continuation that belongs to $\mathsf{p}=\mathsf{0}$ is a stress-free straight fiber with constant polymer mass fraction, speed and temperature (taken from the inflow), i.e.,
\begin{equation*}
\begin{aligned}
\bar{c} = c_0, \qquad u=1, \qquad \sigma=0, \qquad \bar{T}=1. 
\end{aligned}
\end{equation*}
Obviously, it satisfies the prescribed boundary conditions at $z=0$ and $z=1$, since $\mathrm{Dr}_0 = 1$. The change of each continuation parameter has different effects on the form of the solution and hence on finding appropriate initial guesses in the continuation. To navigate through the high-dimensional parameter space from $\mathsf{p}=\mathsf{0}$ to $\mathsf{p}=\mathsf{\underline 1}$ we follow a strategy that consists of two steps:
\begin{enumerate}
\item[(A)] From $\mathsf{p}=\mathsf{0}$ to $\mathsf{p}^A=(1,1,1,0,0,1)$: \\
By changing the parameters $p_1$, $p_2$, $p_\mathrm{Re}$ and $p_\mathrm{Dr}$ from $0$ to $1$ we include mass and heat transfer due to evaporation as well as viscous and drawing effects.
\item[(B)] From $\mathsf{p}^A$ to $\mathsf{p} = \mathsf{\underline{1}}$:\\
By increasing the parameters $p_\mathrm{Fr}$, $p_\mathrm{M}$ from $0$ to $1$ we incorporate gravitational and aerodynamic effects in our solution.
\end{enumerate}
We use a heuristic that seeks for the diagonal continuation path through the parameter space. In both Steps (A) and (B) it turns out that balancing the physical effects strictly leads to the diagonal path such that the parameter space can be reduced to $p_1 = p_2 = p_\mathrm{Re}=p_\mathrm{Dr}$ and $p_\mathrm{Fr}=p_\mathrm{M}$ respectively. The corresponding one-dimensional search is performed with the adaptive step size control as described before.

\subsubsection{Coupling step} Let $\langle \mu(c,T) \rangle_{R^2}^{(i)}$ be the integrated dynamic viscosity of iteration $i$ in the coupling step (cf.\ Algorithm~\ref{sec:numerics_algo1}). We introduce a continuation parameter $p_\mu \in [0,1]$ for this quantity, i.e., we replace the integrated dynamic viscosity in iteration $i$ by the linear combination with the viscosity of the previous iteration $(1-p_\mu)\langle\mu(c,T) \rangle_{R^2}^{(i-1)} + p_\mu\langle\mu(c,T) \rangle_{R^2}^{(i)}$. Thereby, we set $\langle\mu(c,T) \rangle_{R^2}^{(0)} = \mu(\bar{c},\bar{T})R^2$. Again, the corresponding one-dimensional search is performed with the adaptive step size control as described before.

\subsection{Green's function and product integration method}
Consider a boundary value problem of the form
\begin{equation}\label{sec:numerics_eq:diff-trafod}
\begin{aligned}
\partial_x\psi - \frac{1}{r}\partial_r(r\partial_r\psi) &= 0,\\
\partial_r\psi\big|_{r=0} = 0,\qquad \partial_r\psi\big|_{r=1} &= a\psi\big|_{r=1}+b,\qquad \psi\big|_{x=0} = \psi_0,
\end{aligned}
\end{equation}
with $x$ denoting the second component. For constant function $a=const$ the solution can be given analytically in terms of an explicit expression. For non-constant $a$ an implicit solution expression for $\psi$ with Green's function $g$
is found in \cite{arpaci:b:1991, cole:b:2010, ozisik:b:2013}, which reads for constant initial data $\psi_0 = const$,
\begin{align}\label{sec:numerics_eq:psi}
\psi(r,x) &= \psi_0 + 2\pi\int\limits_0^x g(r,x-x') k(x',\psi(1,x'))\,dx',\\\nonumber
\text{with} \quad g(r,z) &= \frac{1}{\pi}\left(1+\sum\limits_{m=1}^\infty\frac{J_0(\beta_mr)}{J_0(\beta_m)}\exp(-\beta_m^2 z)\right), \qquad k(x,y) = a(x)y + b(x),
\end{align}
where $J_i$ denotes the $i$-th Bessel function of the first kind and $\beta_m$ are the ascending zeros of the first Bessel function of the first kind, i.e., $J_1(\beta_m) = 0$. These values are tabulated in literature, see e.g.\ \cite{cole:b:2010}. For $\psi|_{r=1}$, the solution expression yields a Volterra integral equation of second kind
\begin{align}\label{sec:numerics_eq:IntEq}
\psi(1,x) &= \psi_0 + 2\pi\int\limits_0^x g(1,x-x')k(x',\psi(1,x'))dx',\\\nonumber
\text{with} \quad g(1,z) &= \frac{1}{\pi}\left(1+\sum\limits_{m=1}^\infty\exp(-\beta_m^2 z)\right).
\end{align}
The integral kernel $g$ is singular for $z=0$. Numerical integration in the sense of quadrature formulas cannot be applied directly to the integral in (\ref{sec:numerics_eq:IntEq}), because they involve the evaluation of the integrand function at or close to the singularity. Therefore, we use the product integration method, which means substituting the function $k(\cdot,\psi(1,\cdot))$ piecewise by Lagrange interpolation polynomials and employing iterated integration by parts to isolate the singularity of the kernel function $g$. To keep consistency we use quadratic polynomials with nodes corresponding to the Lobatto~IIIa collocation scheme (\ref{sec:numerics_eq:collocation}) and underlying mesh points $x_i$, $i=0,...,N_x$ with mesh size $l_i = x_i-x_{i-1}$
\begin{align*}
\psi(1,x_i) = \psi_0 + 2\pi\sum\limits_{j=1}^{i-1}\int\limits_{x_j}^{x_{j+1}} g(1,x_i-x') \bigg(& k(x_j,\psi_j)\frac{(x'-x_{j+1/2})(x'-x_{j+1})}{(x_j-x_{j+1/2})(x_j-x_{j+1})}\\
& + k(x_{j+1/2},\psi_{j+1/2})\frac{(x'-x_{j})(x'-x_{j+1})}{(x_{j+1/2}-x_{j})(x_{j+1/2}-x_{j+1})} \\
& + k(x_{j+1},\psi_{j+1})\frac{(x'-x_{j})(x'-x_{j+1/2})}{(x_{j+1}-x_{j})(x_{j+1}-x_{j+1/2})}\bigg)dx',
\end{align*}
where
\begin{align*}
x_{j+1/2} &= \frac{1}{2}(x_j+x_{j+1}),\qquad \psi_j=\psi(1,x_j),\\
\psi_{j+1/2} &= \frac{1}{2}(\psi_j+\psi_{j+1}) + \frac{l_j}{8}\left(k(x_j,\psi_j) - k(x_{j+1},\psi_{j+1})\right).
\end{align*}
Applying integration by parts consecutively three times yields a linear system of equations
\begin{align*}
(\mathrm{Id}-2\pi G)\cdot\Psi = \Psi_0 + 2\pi H
\end{align*}
for the unknown vector $\Psi=(\psi_1,...,\psi_N)\in\mathbb{R}^{N_x}$ with matrix $G =(G_{ij})_{ij} \in\mathbb{R}^{{N_x}\times {N_x}}$
\begin{align*}
 G_{ij}= \begin{cases}a_1 A_{i1} + D_1 B_{i1}&, \text{if } j=1 \wedge i > 1,\\
a_j(A_{ij}+ C_{ij-1}) + D_j B_{ij} +  E_{j-1} B_{ij-1}&, \text{if } 1 < j < i \wedge i > 1,\\
E_{i-1} B_{ii-1} + a_i C_{ii-1}&, \text{if } j= i \wedge i > 1,\\
0&,\text{else},
\end{cases}
\end{align*}
as well as right hand side vectors $\Psi_{0}=\psi_0\mathsf{\underline 1}\in\mathbb{R}^{N_x}$ and
\begin{align*}
H = \bigg(b_1 A_{i1} + F_1 B_{i1} + b_i C_{ii-1} + \sum\limits_{j=2}^{i-1}
b_j(A_{ij} + C_{ij-1}) + F_jB_{ij}\bigg)_{i=1,...,N}\in\mathbb{R}^{N_x}.
\end{align*}
Here, we use the abbreviations
\begin{align*}
A_{ij} &= g_{ij}^{(-1)} - \frac{1}{l_j}(g_{ij+1}^{(-2)} + 3g_{ij}^{(-2)}) + \frac{4}{l_j^2}(-g_{ij+1}^{(-3)}+g_{ij}^{(-3)}),\\
B_{ij} &= \frac{4}{l_j}(g_{ij+1}^{(-2)}+g_{ij}^{(-2)}) - \frac{8}{l_j^2}(-g_{ij+1}^{(-3)}+g_{ij}^{(-3)}),\\
C_{ij} &= -g_{ij+1}^{(-1)} - \frac{1}{l_j}(3g_{ij+1}^{(-2)} + g_{ij}^{(-2)}) + \frac{4}{l_j^2}(-g_{ij+1}^{(-3)}+g_{ij}^{(-3)}),\\
D_j &= \frac{1}{2}a_{j+1/2} + \frac{l_j}{8}a_{j+1/2}a_j,\qquad\qquad
E_j = \frac{1}{2}a_{j+1/2} - \frac{l_j}{8}a_{j+1/2}a_{j+1},\\
F_j &= a_{j+1/2}\frac{l_j}{8}(b_j-b_{j+1}) + b_{j+1/2},
\end{align*}
as well as $a_j = a(x_j)$, $b_j = b(x_j)$ and the primitives
\begin{align*}
g_{ij}^{(-\nu)} &= g^{(-\nu)}(1,z_i-z_j),\quad g^{(-\nu)}(r,x) = \int\limits_0^x g^{(-(\nu -1))}(r,x')dx',\quad g^{(0)}(r,x) = g(r,x),
\end{align*}
which can be evaluated analytically. The system matrix is lower triangular, such that the linear system of equations can be solved by forward substitution.

For the computation of the radial profiles we introduce an equidistant grid in $r$-direction with grid points $r_j$, $j= 0,...,N_r$. Then, fixing a point $r_j$, $j < N_r$, and considering the solution expression \eqref{sec:numerics_eq:psi} the solution $\psi(r_j,x_i)$ can be calculated straightforwardly for all $i=0,...,N_x$ by integration using the product integration method analogously. For given profile the calculation of the averaged quantity $\langle \psi \rangle_{R^2}/\pi$ is done by numerical quadrature in $r$-direction employing the standard Lobatto~IIIa collocation method.

In the following we explain the applicability of the general form (\ref{sec:numerics_eq:diff-trafod}) to the two-dimensional advection-diffusion equations (\ref{sec:model_eq:simplified_2d}) of our fiber model. Assuming non-zero coefficients, the equations (\ref{sec:model_eq:simplified_2d}) for polymer mass fraction $c$ and mixture temperature $T$ can be written in similar form
\begin{equation*}
\begin{aligned}
\partial_z c - \lambda_c\frac{1}{r}\partial_{r}(r\partial_r c) &= 0,\qquad
&\partial_z T - \lambda_T\frac{1}{r}\partial_r(r \partial_r T) &= 0,
\end{aligned}
\end{equation*}
with $ \lambda_c = \bar{c}D(\bar{c},\bar{T})/(\epsilon\mathrm{Pe}_Du R^2)$ and $\lambda_T = C(\bar{c},\bar{T})/(\epsilon\mathrm{Pe}_C \rho(\bar{c},\bar{T}) q(\bar{c},\bar{T}) u R^2)$. To obtain the general form \eqref{sec:numerics_eq:diff-trafod} we transform the unknowns $c$ and $T$ by the following rule
\begin{align*}
\psi(r,\Lambda(z)) &= \xi(r,z), \quad \text{ with } \quad \Lambda(z) = \int\limits_0^z \lambda_\xi(z')dz', \quad \xi\in\{c,T\}. 
\end{align*}
The initial conditions become $\psi_0 = c_0$ as well as $\psi_0 = 1$. The coefficient functions are 
\begin{align*}
a(x) = -\mathrm{Pe_D}\frac{R\gamma(c,T)\big|_{r=1}}{\rho(\bar{c},\bar{T}) D(\bar{c},\bar{T})}\bigg|_{z=\Lambda^{-1}(x)},\qquad b(x) = -a(x)c_{ref}(c,T)\big|_{r=1, z=\Lambda^{-1}(x)},
\end{align*}
for the mass fraction equation
and 
\begin{align*}
a(x) &= -\mathrm{Pe_C}\frac{R\alpha}{C(\bar{c},\bar{T})}\bigg|_{z=\Lambda^{-1}(x)},\\
b(x) &= -a(x)T_\star\big|_{z=\Lambda^{-1}(x)} - \mathrm{Pe_C}\frac{R(j(c,T)(\delta(T)-h_d^0(T)))\big|_{r=1}}{C(\bar{c},\bar{T})}\bigg|_{z=\Lambda^{-1}(x)},
\end{align*}
for the temperature equation, respectively. Considering the iteration procedure in Step (2) of Algorithm~\ref{sec:numerics_algo1} the coefficient functions $a$ and $b$ are evaluated at $c|_{r=1}$, $\bar{c}$ and $T|_{r=1}$, $\bar{T}$ of the previous iteration step, i.e., the boundary conditions are treated in a semi-implicit way, such that the linear Robin-type character is preserved. This is certainly the key point for the embedding into the solution framework of Green's functions and product integration method that ensures the desired numerical efficiency.

\begin{remark}[Grid resolution]
The mesh points in $x$-direction are chosen in such a way that they correspond to the underlying mesh points from the collocation method, i.e., $x_i= \Lambda(z_i)$. If boundary layers at the nozzle are expected, the mesh is refined there. For the equidistant points in $r$-direction we set $N_r= 101$.
\end{remark}
\begin{remark}[Stability for stiff problems]
Since the product integration method inherits its stability properties from the underlying quadrature method, our method based on the Lobatto~IIIa collocation method is A-stable but not L-stable. Thus, it can lead to oscillations for very stiff problems. If oscillations occur, we switch our product integration method to a method with underlying implicit Euler scheme, lowering the order of convergence but increasing the stability.
\end{remark} 

\subsection{Iterative coupling of fiber and airflow computations} 
The fiber-air-interaction is realized by a weak coupling algorithm which iterates between the described fiber computation and the airflow simulation. This procedure allows the combination of our self-implemented code for the fiber dynamics and a commercial software for the air dynamics and was successfully used in studies on glass wool manufacturing \cite{arne:p:2011}. The two-way-coupling is based on a homogenization concept and done in the given framework by the help of respective momentum, heat and mass exchange models (drag forces as well as heat and mass sources) that obey the principle that action equals reaction, cf.\ \cite{cibis:p:2014, cibis:p:2017}.

We employ ANSYS Fluent\footnote{For details on the commercial software ANSYS Fluent, its models and solvers we refer to \texttt{http://www.ansys.com}.}, a finite-volume-based software, that contains the broad physical modeling capabilities to describe airflow, heat and mass transfer for the dry spinning process. The airflow is modeled using stationary compressible Navier-Stokes equations, supplemented with the Fourier law for heat conduction as well as thermal and caloric equations of state for an ideal gas. The exchange models are incorporated by UDFs (user defined functions). Considering $M$ representative fibers, $M\in \mathbb{N}$, and neglecting fiber-fiber-interactions, the fibers are computed in parallel on basis of Algorithm~\ref{sec:numerics_algo1} which is implemented in MATLAB. Because of the different discretizations of airflow and fibers, the exchange of airflow and fiber data between the solvers requires interpolation and averaging. For the air associated exchange models the airflow data are linearly interpolated on the fiber grid used in the collocation method. Non-uniform profiles of air velocity, air temperature and diluent mass fraction in the air generally induce air associated exchange models (drag forces as well as heat and mass sources) being different for each fiber. For the fiber associated exchange models entering the finite volume scheme, the information of all fibers is spatially averaged over each cell volume. Since the interpolation and averaging strategies are qualitatively different, the conservation principles are only ensured for very fine resolutions. Moreover, note that due to the underlying finite volume approach the computed airflow data are cell-averaged quantities. In this context the numerical cell-averaging can be interpreted as homogenization strategy for the exchange models that is necessary to avoid the occurrence of singularities and the failure of the two-way-coupling. For details we refer to \cite{cibis:p:2014} for the two-way-coupling within an asymptotic modeling framework, to \cite{cibis:p:2017} for the homogenization and to \cite{arne:p:2011} for the algorithmic procedure.

Let $\mathcal{S}$ and $\mathcal{S}_\star$ be the algorithmic simulation procedures to obtain the quantities $\Psi=(\Psi_k)$, $k \in \{1,...,M\}$ of the $M$ fibers and the quantities $\Psi_\star$ of the airflow, respectively. Then, the basic coupling algorithm has the following form:
\begin{algorithm}[Fiber-airflow coupling]~
\begin{enumerate}
\item[(1)] Perform an airflow simulation $\mathcal{S}_\star$ without fibers to obtain $\Psi_\star^{(1)}$.
\item[(2)] For $i \geq 1$, calculate $\Psi_k^{(i)} = \mathcal{S}(\Psi_\star^{(i)})$ for all fibers $k=1,...,M$ and afterwards compute the new airflow data $\Psi_\star^{(i+1)} = \mathcal{S}_\star(\Psi^{(i)})$, as long as $\lVert \Psi_\star^{(i)} - \Psi_\star^{(i+1)}\rVert > tol$ for given tolerance $tol$.
\end{enumerate}
\end{algorithm}

\setcounter{equation}{0} \setcounter{figure}{0} \setcounter{table}{0}
\section{Approximation quality of asymptotic model and performance of algorithm} \label{sec:performance}
\begin{table}[t]
\begin{minipage}[t]{0.49\textwidth}
\begin{center}
\begin{small}
\vspace*{-2.25cm}
\begin{tabular}{| l r@{ = }l |}
\hline
\multicolumn{3}{|l|}{\textbf{Process Parameters}}\\
\hline
fiber length & $L$ & $0.2$ m\\
nozzle radius & $R_0$ & $4.07\cdot10^{-4}$ m\\
speed at nozzle & $u_0$ & $ 5\cdot10^{-2}$ m/s\\
temperature at nozzle & $T_0$ & $ 363.15$ K\\
polym. mass fract. at nozzle & $c_0$ &$ 0.6$\\
temperature of surrounding & $T_\star$ & $350$ K\\
diluent density of surrounding   &  \hspace*{-0.2cm} $\rho_{d,\star}$ & $0$ kg/m$^3$\\
drag force & \hspace*{-0.2cm} $f_{air}$ & $0$ N/m\\
\multicolumn{3}{|l|}{}\\
\multicolumn{3}{|l|}{}\\
\hline
\end{tabular}\end{small}
\end{center}
\end{minipage}
\hfill
\begin{minipage}[t]{0.49\textwidth}
\begin{center}
\begin{small}
\begin{tabular}{| l r@{ = }l |}
\hline
\multicolumn{3}{|l|}{\textbf{Physical Parameters}}\\
\hline
polymer density & $\rho_p^0$& $1350$ kg/m$^3$\\
diluent density & $\rho_d^0 $& $1000$ kg/m$^3$\\
diluent enthalpy & $h_d^0$& $2\cdot10^5$ m$^2$/s$^2$\\
specific heat capacity & $q$ & $2750$ m$^2$/(s$^2$K)\\
dynamic viscosity & $\mu $&$200$ Pa\,s\\
mass transfer coeff. & $\gamma$ & $0.3$ kg/(m$^2$s)\\
heat transfer coeff. & $\alpha$ & $2000$ W/(m$^2$K)\\
evap. enthalpy diluent & $\delta$ &$2.35\cdot10^6$ J/kg\\
diffusion coefficient & $D$&$3.5\cdot 10^{-10}$ m$^2$/s\\
thermal conductivity & $C$&$0.4$ W/(m K)\\
\hline
\end{tabular}
\end{small}
\end{center}
\end{minipage}\\~\\~\\
\caption{Overview over process and physical parameters used for the performance analysis, yielding $(\epsilon, \mathrm{Re}, \mathrm{Fr}, \mathrm{Pe_D}, \mathrm{Pe_C}) = (2\cdot 10^{-3}, 5.92\cdot 10^{-2}, 3.57\cdot 10^{-2}, 5.81\cdot 10^{4}, 1.65\cdot 10^{2}) $.}\label{sec:performance_table:param}
\end{table}
\begin{figure}[t]
\centering
\includegraphics{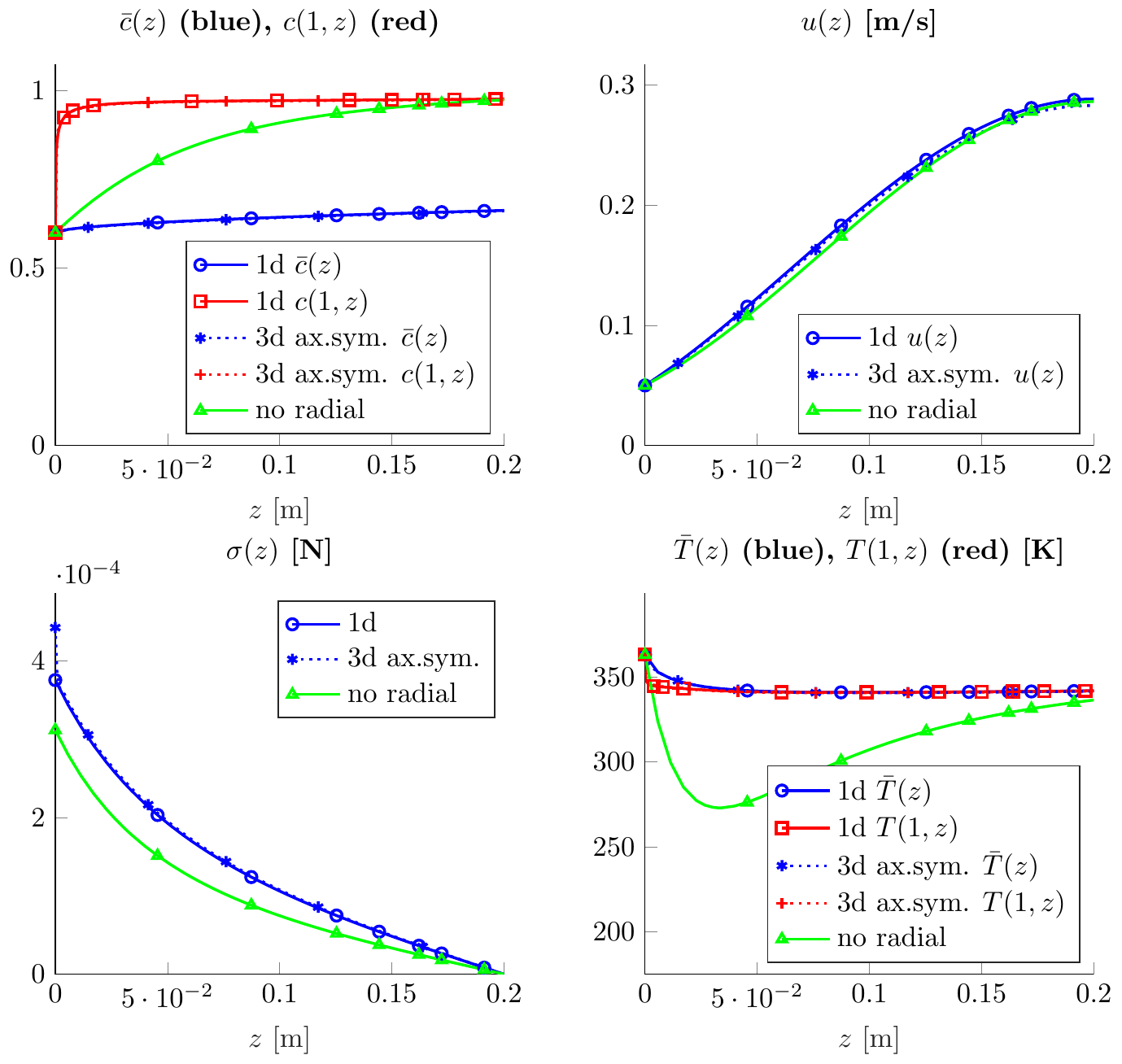}
\caption{One-dimensional fiber solution obtained with and without radial information (\textit{solid lines}) in comparison to the reference solution (\textit{dotted lines}).}\label{sec:performance_fig:1d}
\end{figure}
\begin{figure}[t]
\begin{minipage}[c]{0.49\textwidth}
	\centering
	\includegraphics{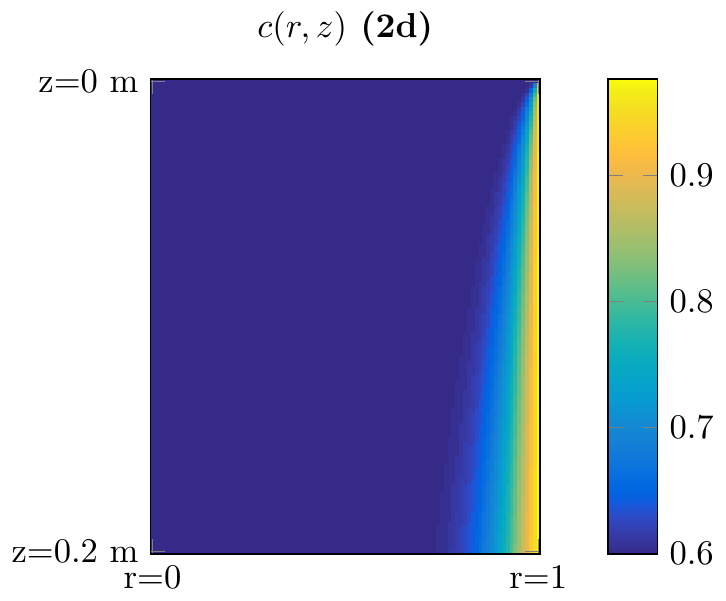}
\end{minipage}\hfill
\begin{minipage}[c]{0.49\textwidth}
	\centering
	\includegraphics{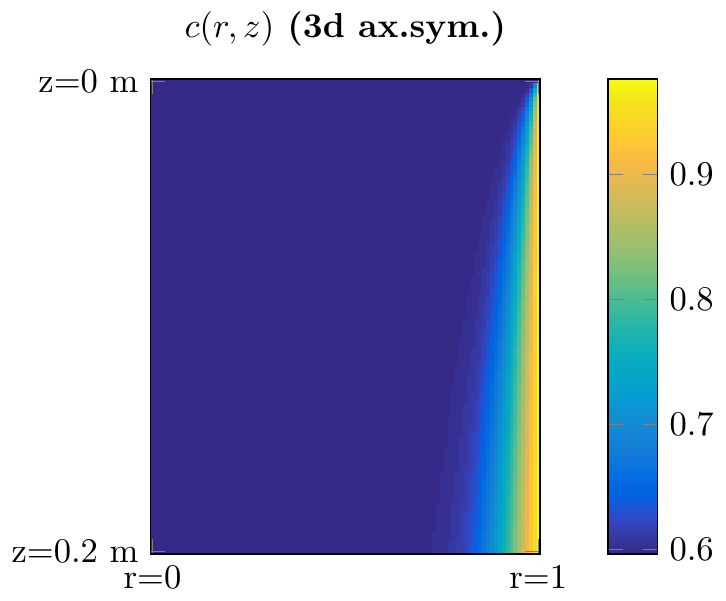}
\end{minipage}
\begin{minipage}[c]{0.49\textwidth}
	\centering
	\includegraphics{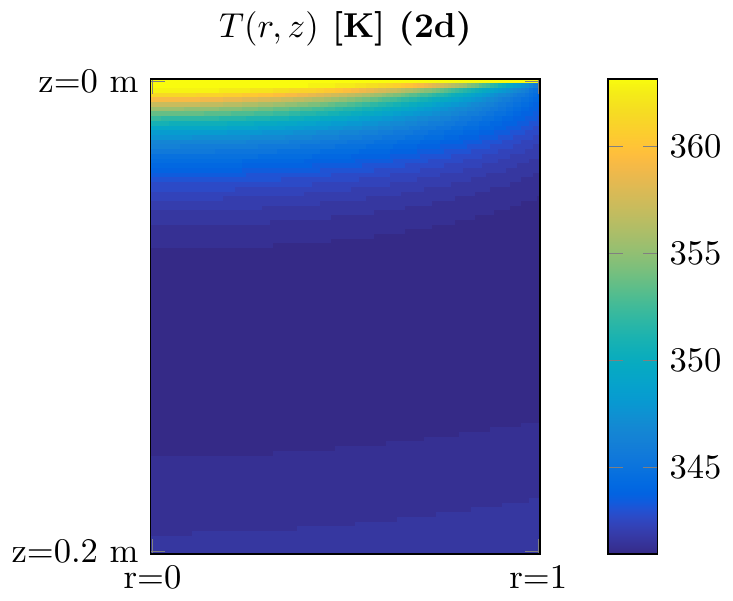}
\end{minipage}\hfill
\begin{minipage}[c]{0.49\textwidth}
	\centering
	\includegraphics{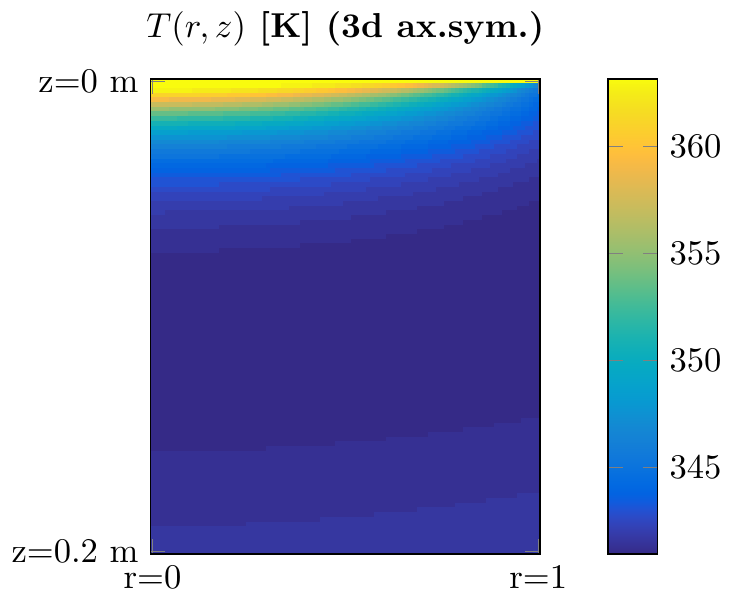}
\end{minipage}
\caption{Polymer mass fraction and temperature profiles as solution of the one-two-dimensional fiber model (\textit{left}) and the reference model (\textit{right}).}\label{sec:performance_fig:profiles}
\end{figure}
\begin{table}[t]
\centering
\begin{tabular}{ c | c | c | c }
$\bar{c}$ & $u$ & $\sigma$ & $\bar{T}$\\ \hline\hline
$1.8064\cdot10^{-3}$  &  $1.5301\cdot10^{-2}$  &  $4.9611\cdot10^{-2}$  &  $6.4668\cdot10^{-4}$
\\ 
\multicolumn{1}{c}{}\\
 $c|_{r=1}$ & $T|_{r=1}$ & $c$ & $T$\\ \hline\hline
$4.3393\cdot10^{-3}$  &  $1.1876\cdot10^{-3}$   &   $4.1332\cdot10^{-3}$  &  $7.2008\cdot10^{-4}$
\\
\multicolumn{1}{c}{}\\
\end{tabular}
\caption{Relative L$^2$-errors between one-two-dimensional solution and reference solution, cf.\ Figs.~\ref{sec:performance_fig:1d} and \ref{sec:performance_fig:profiles}.}\label{sec:performance_table:l2errors}
\end{table}
To study the approximation quality of the proposed dimensionally reduced fiber model (System~\ref{sec:model_eq:coupled_sys_simpl}) and the performance of the numerical solution algorithm, we compare our developed procedure for a single fiber with the solution of the original three-dimensional problem in System \ref{sec:model_eq:3d_model}. As the direct numerical simulation of a three-dimensional thin fiber immersed in an airflow in a spinning chamber is computationally extremely demanding due to the required high resolution, we limit ourselves to a three-dimensional axis-symmetric setting of a fiber without any aerodynamic forces and with stress-free end. This implies the boundary conditions $\boldsymbol{\Sigma}\cdot \boldsymbol{\nu}=\mathbf{0}$ on $\Gamma_{\mathrm{fr}}\cup \Gamma_{\mathrm{out}}$ in System~\ref{sec:model_eq:3d_model} and $f_{air}=0$, $\sigma(1)=0$ in System~\ref{sec:model_eq:coupled_sys_simpl}. Moreover, we consider a dry surrounding, i.e., $\rho_{d,\star}=0$ (implying $c_{ref}=1$), and take all physical parameters as well as the mass and heat sources as constant. 

We refer to the three-dimensional problem as reference and solve it with FEniCS\footnote{For details on the open-source finite element code FEniCS see \texttt{http://www.fenicsproject.org}.}, an open-source finite element code with Python interface, which basically needs a weak formulation of the partial differential model equations. To overcome the need of re-meshing while solving the reference, we transform the weak form of System \ref{sec:model_eq:3d_model} to the fixed domain $\tilde{\Omega}_{cut} = (0,1)\times(0,L)$. The resulting ordinary differential equation for the fiber radius function $R$ on the domain boundary $\{1\}\times(0,L)$ and the balance equations are solved by iterations. The domain $\tilde{\Omega}_{cut}$ is meshed with a triangular grid, which is refined at the nozzle $\Gamma_{\mathrm{in}}$ due to expected boundary layers. The nodes of the triangles are $(r_i,z_j)$ with $r_i = i/200$, $z_j = L(j/40)^{1.5}$ for $i=0,...,200$, $j=0,...,40$, which means 16000 triangular elements and 8241 nodes in total.
Algorithm~\ref{sec:numerics_algo1} for the dimensionally reduced fiber model is implemented in MATLAB version R2016b, where the boundary value solver \texttt{bvp4c.m} is used with the default values. To achieve comparability the reference solution is interpolated on the mesh of the solution of System~\ref{sec:model_eq:coupled_sys_simpl}, i.e., the mesh with the collocation points used in the boundary value solver (\texttt{bvp4c.m}) in $z$-direction and 101 equidistant points in $r$-direction. As stopping criterion for the iteration in the coupling step of Algorithm~\ref{sec:numerics_algo1} we use $tol = 10^{-6}$. The forthcoming numerical simulations are performed on an Intel Core i7-6700 CPU (4 cores, 8 threads) and 16 GBytes of RAM.

As an example we consider a set-up with moderate thinning and evaporation effects. The physical and process parameters are particularly chosen as listed in Table \ref{sec:performance_table:param}, implying $\epsilon = 2\cdot 10^{-3}$, Reynolds number $\mathrm{Re} = 5.92\cdot 10^{-2}$, Froude number $\mathrm{Fr} = 3.57\cdot10^{-2}$ as well as Peclet numbers $\mathrm{Pe_D} = 5.81\cdot 10^4$ and $\mathrm{Pe_C} = 1.65\cdot 10^2$. 
Figure~\ref{sec:performance_fig:1d} shows the solution curves obtained from the dimensionally reduced fiber model of System~\ref{sec:model_eq:coupled_sys_simpl}. This solution we oppose the solution obtained without radial information, which means exiting Algorithm~\ref{sec:numerics_algo1} after Step (1), and the three-dimensional axis-symmetric solution of the reference. For the latter the quantities of interest $\bar{c}$, $\bar{T}$, $u$ and $\sigma$ are calculated a posteriori according to the definitions in Section~\ref{sec:model}. The corresponding two-dimensional profiles for polymer mass fraction $c$ and temperature $T$ can be seen in Fig.~\ref{sec:performance_fig:profiles}. We draw the following conclusions from the plots: As expected the coupling of the one-dimensional model equations with the two-dimensional equations for $c$ and $T$ is absolutely essential, since the solution without that information differs extremely from the reference solution, especially concerning the averaged quantities $\bar{c}$ and $\bar{T}$. This point is emphasized by the non-uniform profiles in Fig.~\ref{sec:performance_fig:profiles}. The solution of the one-two-dimensional fiber model approximates the reference solution very accurately and takes the important radial effects into account. The relative L$^2$-errors between the single components of the solutions, as listed in Table \ref{sec:performance_table:l2errors}, verify the utilization of our model framework with regard to the good approximation quality. In addition, the difference in the computation times is crucial. While the computation of the three-dimensional axis-symmetric reference solution takes around 4228 seconds, which means more than 1.17 hours, the computation of the reduced model takes only around 43 seconds (around 4 seconds for the initialization step (Step (1)) and 39 seconds for the coupling step (Step (2)) of Algorithm~\ref{sec:numerics_algo1}). The good performance is owed to our embedding of the two-dimensional advection-diffusion equations into the solution framework of Green's functions and product integration method. In fact, the achieved efficiency is essential when dealing with more than one fiber and considering fiber-airflow-interactions in a spinning column via two-way coupling. So, our proposed procedure builds obviously a good basis for the simulation of industrial dry spinning processes.
\begin{remark}[Temperature behavior]
In this academic example the behavior of the fiber temperature $T$ might be surprising (cf. Fig.~\ref{sec:performance_fig:profiles}). Through solvent evaporation associated heat transfer $T$ drops below $T_\star$ shortly after the nozzle. Due to the relative temperature between fiber and surrounding being negative, i.e., $T-T_\star < 0$, the fiber is heated up by the surrounding towards the fiber end, when mass transfer effects are not dominating the heat transfer anymore. This behavior is also observed in the industrial case, when only an one-way coupling with the surrounding airflow is considered (cf. Fig.~\ref{sec:ex_fig:it1_30}). However, regarding full two-way coupled airflow this behavior does not occur.
\end{remark}

\setcounter{equation}{0} \setcounter{figure}{0} \setcounter{table}{0}
\section{Dry spinning with two-way-coupled airflow}\label{sec:example}
In industrial dry spinning applications we are faced with multiple jets being extruded from the spinneret into the spinning column. The properties of the spun fibers are essentially determined by the fiber-airflow-interactions. Hence, we aim for a simulation of the dry spinning with two-way-coupled airflow. Until now the existing literature solely considers fibers immersed in an airflow that is assumed to be unperturbed. In \cite{sano:p:2001}, for example, a cellulose acetate (CA)-acetone mixture is dry spun in a spinning column with counter-current airflow. The respective airflow quantities are measured and fixed, they are not affected by the fiber dynamics, and some characteristics for a single fiber are computed. Consequently, we cannot validate our simulation results, but we check the reasonableness of our model by comparing the intermediate results obtained for a single fiber after the first iteration of our fiber-air coupling algorithm which corresponds to a one-way-coupling (as in \cite{sano:p:2001}). Before we present our simulation outcome, we give a description of the setup and all models employed to close our fiber dry spinning model.

\subsection{Setup}
\begin{table}[t]
\begin{minipage}[c]{0.45\textwidth}
\begin{center}
\begin{small}
\begin{tabular}{| l r@{ = }l |}
\hline
\multicolumn{3}{|l|}{\textbf{Process Parameters}}\\
\hline
device height & $H$ & $5.1$ m\\
speed at nozzle & $u_0$ &$5$ m/s\\
temperature at nozzle & $T_0$ & $ 348.15$ K\\
polym. mass fract. at nozzle & $c_0$ & $0.2882$\\
take up speed at bottom & $u_1$& $10$ m/s\\
total number of fibers & $M$& $60$\\
\multicolumn{3}{|l|}{}\\
\multicolumn{3}{|l|}{}\\
\hline
\end{tabular}
\end{small}
\end{center}
\end{minipage}
\hfill
\begin{minipage}[c]{0.52\textwidth}
\begin{center}
\begin{small}
\begin{tabular}{| l r@{ = }l |}
\hline
\multicolumn{3}{|l|}{\textbf{Physical Parameters}}\\
\hline
CA density & $\rho_p^0$ &$1300$ kg/m$^3$\rule{0pt}{2.6ex}\\
acetone density & $\rho_d^0 $&$ 767$ kg/m$^3$\\
CA spec. heat cap. & $q_{p}^0$ &$1600$ J/(kgK)\\
acetone spec. heat cap. & $q_{d}^0 $&$2160$ J/(kgK)\\
molec. weight acetone & $M_d $&$5.808\cdot 10^{-2}$ kg/mol\\
molec. weight air & $M_\star$&$2.897\cdot 10^{-3}$ kg/mol\\
molar volume acetone & $V_d $&$6.60\cdot 10^{-5}$ m$^3$/mol\\
molar volume air & $V_\star$&$2.01\cdot 10^{-5}$ m$^3$/mol\\
\hline
\end{tabular}
\end{small}
\end{center}
\end{minipage}\\~\\~\\
\caption{Overview over process and physical parameters in the industrial dry spinning setup \cite{sano:p:2001}, yielding $(\epsilon, \mathrm{Re}, \mathrm{Fr}, \mathrm{Pe_D}, \mathrm{Pe_C}) = (6.16\cdot 10^{-5},1.79\cdot 10^{3}, 7.07\cdot 10^{-1}, 2.08\cdot 10^{7}, 1.37\cdot 10^{4})$.}\label{sec:ex_table:param}
\end{table}
\begin{figure}[t]
\centering
\includegraphics[height=4.5cm]{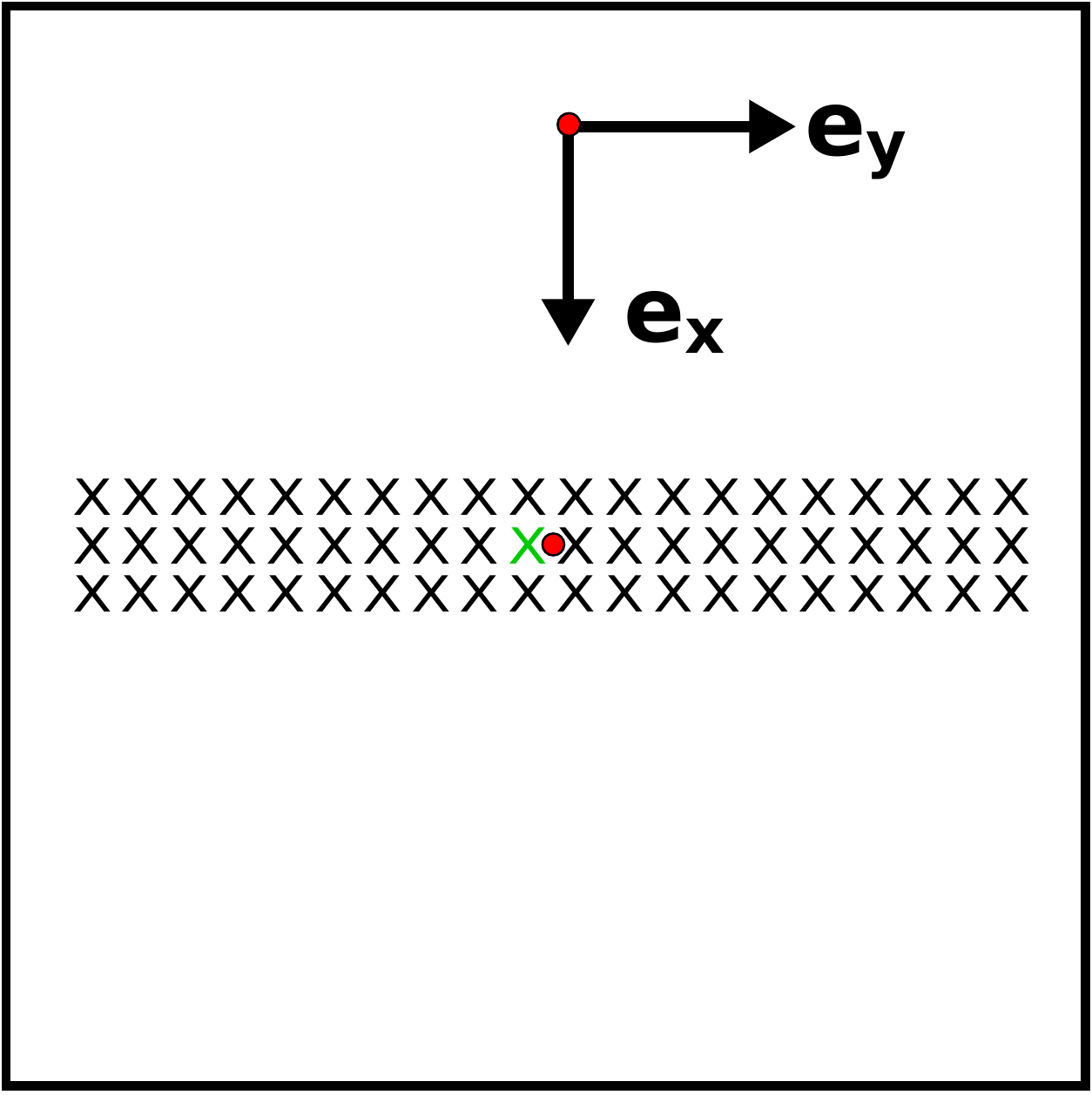}
\caption{Dry spinning setup, sketch of the holes \textit{(X)} in the spinneret. All proceeding simulation results are shown for the fiber extruded from the central position \textit{(green X)}.}\label{sec:ex_fig:spinneret}
\end{figure}
We consider the dry spinning setup of \cite{sano:p:2001} where a CA-acetone mixture is pumped from the spinneret through a row of holes into a spinning column with counter-current airflow (cf.\ Fig.~\ref{sec:intro_fig:apparatus}). For a larger emergence of the effects of fiber-air interaction, we look at sixty holes (instead of twenty as in \cite{sano:p:2001}), which are ordered in three parallel rows with twenty holes each, as sketched in Fig.~\ref{sec:ex_fig:spinneret}. We consider the fibers to be spun in the $y$-$z$-plane and the ones of the middle row particularly to lie in $x=0$. At the column bottom all fibers are drawn down by a take up roller. Concerning the airflow, we assume dry air at the inlet, i.e., the inflow contains no diluent. The absolute air inflow velocity (in $y$-direction) at the bottom pipe is $\|\mathbf{v}_\star\| = 0.22$~m/s and the air inflow temperature is $T_\star=330$~K. At the nozzle the CA-acetone mixture forms a drop, from which a jet of initial diameter usually larger than the initial nozzle diameter leaves due to the Barus effect. Since our model does not take this effect into account, we set $R_0$ according to the initial speed $u_0$ and the initial cumulated mixture flux $A_0u_0 = 3.1\cdot10^{-7}$~m$^3$/s,  $A_0=R_0^2\pi$, as prescribed in \cite{sano:p:2001}. The remaining process parameters are listed in Table~\ref{sec:ex_table:param}.

\subsection{Closing models}\label{subsec:closing}
The physical material parameters for the polymer CA and the diluent acetone are taken from \cite{sano:p:2001, gou:p:2004}, see Table~\ref{sec:ex_table:param}. The additional models (exchange models as well as rheological models) that we use to close our dimensionally reduced fiber model (System \ref{sec:model_eq:coupled_sys_simpl}) are briefly summarized in this subsection (cf.\ Remark~\ref{sec:model_remark}). Note that to distinguish the fiber quantities from the airflow quantities all airflow-associated fields are labeled here with the index $_\star$ as before. In particular, $\mathbf{v}_\star$ [m/s] denotes the velocity, $p_\star$ [Pa] the pressure, $\rho_\star$ [kg/m$^3$] the density, $\mu_\star$ [Pas] the dynamic viscosity, $q_{\star}$ [J/(kgK)] the specific heat capacity and $k_\star$ [J/(msK)] the thermal conductivity of the air.

\subsubsection{Drag forces}\label{subsec:drag}
The aerodynamic drag forces $f_{air}$ are described by means of a dimensionless air drag function $\mathbf{F}$ that depends on fiber tangent $\boldsymbol{\tau}$ and relative velocity between airflow and fiber,
\begin{align*}
f_{air} &= \frac{\mu^2_\star}{2R\rho_\star} \mathbf{F}\bigg(\boldsymbol{\tau},\frac{2R\rho_\star}{\mu_\star}(\mathbf{v}_\star - u\boldsymbol{\tau})\bigg)\cdot \boldsymbol{\tau},\\
\mathbf{F}(\boldsymbol{\tau},\mathbf{w}) &= w_\nu r_\nu(w_\nu)\boldsymbol{\nu} + w_\tau r_\tau(w_\nu)\boldsymbol{\tau}.
\end{align*}
In particular, $\mathbf{F}$ is expressed in terms of the tangential $w_\tau = \mathbf{w}\cdot\boldsymbol{\tau}$ and normal velocity components $w_\nu=\sqrt{\mathbf{w}\cdot\mathbf{w}-w_\tau^2}$ with normal vector $\boldsymbol{\nu} = (\mathbf{w}-w_\tau\boldsymbol{\tau})/w_\nu$. We use the regularized air resistance coefficients $r_\nu$, $r_\tau$ given in \cite{marheineke:p:2009b}
\begin{align*}
r_\nu(w_\nu) &= \begin{cases}
\sum_{j=0}^3 q_{\nu,j}w_\nu^j,\quad & w_\nu < w_0,\\
\frac{4\pi}{S(w_\nu)}\big(1-\frac{S^2(w_\nu)-S(w_\nu)/2+5/16}{32S(w_\nu)}w_\nu^2\big),\qquad \quad &w_0 \leq w_\nu < w_1,\\
w_\nu\exp\big(\sum_{j=0}^3 p_{\nu,j}\log^j(w_\nu)\big), &w_1 \leq w_\nu \leq w_2,\\
2\sqrt{w_\nu}+0.5w_\nu, &w_2 < w_\nu,
\end{cases}\\
r_\tau(w_\nu) &= \begin{cases}
\sum_{j=0}^3 q_{\tau,j}w_\nu^j,\quad &w_\nu < w_0,\\
\frac{4\pi}{(2S(w_\nu)-1)}\big(1-\frac{2S^2(w_\nu)-2S(w_\nu)+1}{16(2S(w_\nu)-1)}w_\nu^2\big),\qquad \, &w_0 \leq w_\nu < w_1,\\
w_\nu\exp\big(\sum_{j=0}^3 p_{\tau,j}\log^j(w_\nu)\big), &w_1 \leq w_\nu \leq w_2,\\
2\sqrt{w_\nu}, &w_2 < w_\nu,
\end{cases}
\end{align*}
with $S(w_\nu) = 2.0022 - \log(w_\nu)$. This model is composed of asymptotic Oseen theory, Taylor heuristic and simulations where the transition points $w_1 = 0.1$ and $w_2 = 100$ are estimated from a least-square approximation of experimental and numerical data. For tangential incident flow situations ($w_\nu \rightarrow 0$) a regularization employs the Stokes theory, yielding the Stokes limits
\begin{align*}
r_\nu^S = \frac{4\pi}{\log(4\varepsilon^{-1})}-\frac{\pi}{\log^2(4\varepsilon^{-1})},\qquad r_\tau^S = \frac{2\pi}{\log(4\varepsilon^{-1})}+\frac{\pi}{2\log^2(4\varepsilon^{-1})},
\end{align*}
as well as the transition point $w_0 =2\exp\big(2.0022-{4\pi}/{r_\nu^S}\big)$ with the regularization parameter $\varepsilon = 3.5\cdot 10^{-2}$. The parameters $p_{k,j}$ and $q_{k,j}$ ($k \in \{\nu,\tau\}$, $j\in\{0,1,2,3\}$) ensure smoothness,
\begin{alignat*}{6}
p_{\nu,0} &= 1.6911, \qquad &p_{\nu,1} &= -6.7222\cdot 10^{-1},\qquad
&p_{\nu,2} &= 3.3287\cdot 10^{-2},\qquad &p_{\nu,3}&=3.5015\cdot 10^{-3},\\
p_{\tau,0} &= 1.1552,\qquad &p_{\tau,1} &= -6.8479\cdot 10^{-1},\qquad
&p_{\tau,2} &= 1.4884\cdot 10^{-2},\qquad &p_{\tau,3}&=7.4966\cdot 10^{-4},
\end{alignat*}
\begin{align*}
q_{k,0} &= r_k^S,\qquad q_{k,1} = 0,\qquad q_{k,2} = \frac{3r_k(w_0)-w_0r_k'(w_0)-3r_k^S}{w_0^2},\\
 q_{k,3} &= \frac{-2r_k(w_0)+w_0r_k'(w_0)+2r_k^S}{w_0^3}.
\end{align*}

\subsubsection{Heat transfer}
The model for the heat transfer coefficient $\alpha$ is given by the following relation to the dimensionless Nusselt number $\mathrm{Nu}$ that is a function of the Reynolds numbers $\mathrm{Re}_\tau$ and $\mathrm{Re}_w$ (ratios of inertial to viscous forces with respect to the relative velocity between air and fiber in tangential direction and in total, respectively) as well as the Prandtl number $\mathrm{Pr}$ (ratio of momentum to thermal diffusivities of the air)
\begin{align*}
\alpha &= \frac{k_\star}{2R}\mathrm{Nu}(\mathrm{Re}_\tau,\mathrm{Re}_w,\mathrm{Pr}),\\
\mathrm{Re}_\tau &= \frac{2R\rho_\star ( \mathbf{v_\star} - u\boldsymbol{\tau})\cdot \boldsymbol{\tau} }{\mu_\star},\qquad \mathrm{Re}_w = \frac{2R\rho_\star\lVert \mathbf{v_\star} - u\boldsymbol{\tau}\rVert }{\mu_\star},\qquad \mathrm{Pr} = \frac{\mu_\star q_{\star}}{k_\star}.
\end{align*}
Note that $\mathrm{Re}_\tau$, $\mathrm{Re}_w$ and $\mathrm{Pr}$ are here no global constants, but scalar fields depending on the airflow quantities varying in space.  In particular, $\mathrm{Re}_\tau=w_\tau$ of Sec.~\ref{subsec:drag}. The model used for the Nusselt number is heuristic
\begin{align*}
\mathrm{Nu}(\mathrm{Re}_\tau,\mathrm{Re}_w,\mathrm{Pr}) = (1-0.5\,{A}(\mathrm{Re}_\tau, \mathrm{Re}_w))
\begin{cases}
{n_1}(\mathrm{Re}_w,\mathrm{Pr}),\qquad &\mathrm{Re}_w\mathrm{Pr}\geq 7.3\cdot 10^{-5},\\
{n_2}(\mathrm{Re}_w,\mathrm{Pr}),\qquad & \mathrm{Re}_w\mathrm{Pr} < 7.3\cdot 10^{-5}.
 \end{cases}
\end{align*}
It goes back to  \cite{sucker:p:1976} where originally a stationary perpendicular laminar flow situation around a cylinder for $\mathrm{Re}_w\mathrm{Pr} \geq 7.3\cdot 10^{-5}$ was studied and has been extended to ensure a smooth transition to the limit value $\mathrm{Nu}=0.1$ for vanishing $\mathrm{Re}_w\mathrm{Pr}\rightarrow 0$,
\begin{align*}
{n_1}(\mathrm{Re}_w,\mathrm{Pr}) &= 0.462(\mathrm{Re}_w\mathrm{Pr})^{0.1} + f(\mathrm{Pr})\frac{(\mathrm{Re}_w\mathrm{Pr})^{0.7}}{1+2.79(\mathrm{Re}_w\mathrm{Pr})^{0.2}},\\
{n_2}(\mathrm{Re}_w,\mathrm{Pr}) &= m_1(\mathrm{Pr})(\mathrm{Re}_w\mathrm{Pr})^3 + m_2(\mathrm{Pr})(\mathrm{Re}_w\mathrm{Pr})^2 + 0.1,
\end{align*}
with the coefficients
\begin{align*}
m_1(\mathrm{Pr}) &= -3.5636\cdot 10^{11} - 3.1380\cdot 10^9 f(\mathrm{Pr}),\qquad m_2(\mathrm{Pr}) = 4.0694\cdot 10^7 + 3.9768\cdot 10^5 f(\mathrm{Pr}),\\
f(\mathrm{Pr}) &= \frac{2.5}{(1 + (1.25\mathrm{Pr}^{1/6})^{2.5})^{0.4}}.
\end{align*}
The incorporation of the function  ${A}$ accounts for varying incident flow directions. It is mainly the squared cosine of the angle of attack, which is regularized to ensure smoothness for a tangential incident flow,
\begin{align*}
{A}(\mathrm{Re}_\tau, \mathrm{Re}_w) =
\begin{cases}
(\mathrm{Re}_\tau \mathrm{Re}_w^{-1})^2 , \qquad &\mathrm{Re}_w \geq \varepsilon,\\
\left( 1- (\mathrm{Re}_w\varepsilon^{-1})^2 \right)^2 + \left(3 - 2(\mathrm{Re}_w\varepsilon^{-1})^2 \right)(\mathrm{Re}_\tau \mathrm{Re}_w\varepsilon^{-2})^2, \qquad &\mathrm{Re}_w < \varepsilon,
\end{cases}
\end{align*}
with regularization parameter $\varepsilon = 10^{-7}$.

\subsubsection{Mass transfer and referential mass fraction} 
The mass fraction associated transfer coefficient $\gamma$ and the referential polymer mass fraction $c_{ref} $ 
\begin{align*}
\gamma = \beta \varrho, \qquad c_{ref} = 1-\frac{\rho_{d,\star}}{\varrho}
\end{align*}
are expressed in terms of the mass transfer coefficient $\beta$, the mass fraction-scaled diluent density in the air at the fiber surface $\varrho$, and the diluent density in the air away from the fiber $\rho_{d,\star}$ (cf.\ Sec.~\ref{sec:model}).  Making use of the analogy of mass transfer and heat transfer the model for $\beta$ by \cite{sucker:p:1976} is of similar form as that for $\alpha$. The Nusselt number $\mathrm{Nu}$ is changed to the Sherwood number $\mathrm{Sh}$ and the Prandtl number $\mathrm{Pr}$ is replaced by the Schmidt number $\mathrm{Sc}$ (ratio of the viscous diffusion rate to the mass diffusion rate) 
\begin{align*}
\beta &= \frac{D_{d,\star}}{2R}\mathrm{Sh}(\mathrm{Re}_\tau,\mathrm{Re}_w,\mathrm{Sc}),\\
 \mathrm{Sc} &= \frac{\mu_\star}{\rho_\star D_{d,\star}}, \qquad 
 \mathrm{Sh}(\mathrm{Re}_\tau,\mathrm{Re}_w, \mathrm{Sc}) = \mathrm{Nu}(\mathrm{Re}_\tau,\mathrm{Re}_w,\mathrm{Sc}),
\end{align*}
with the diffusion coefficient of the diluent in air $D_{d,\star}$ [m$^2$/s]. The coefficient is modeled as \cite{fuller:p:1966}
\begin{align*}
D_{d,\star} = E \frac{T_\star^{1.75}}{p_\star(V_d^{1/3} + V_\star^{1/3})^2}\left(\frac{1}{M_d} + \frac{1}{M_\star}\right)^{0.5},
\end{align*}
with constant $E = 1.01325\cdot 10^{-7.5}$ m$^4$kg$^{1/2}$Pa/(K$^{7/4}$mol$^{7/6}$s) as well as $M_d$ [kg/mol], $M_\star$ [kg/mol] denoting the molecular weights and $V_d$ [m$^3$/mol], $V_\star$ [m$^3$/mol] denoting the molar volumes of diluent and air, respectively. Concerning the diluent density in the air, away from the fiber it is given by $\rho_{d,\star} = \rho_\star c_{d,\star}$ with the diluent mass fraction in air $c_{d,\star}$. At the fiber surface we employ a model for the mass fraction-scaled diluent density $\varrho$ that is based on \cite{bercea:p:2009}  
\begin{align*}
\varrho (c, T) &= \frac{M_d }{R {T}}\,p_{vap}( T)\, \frac{\rho(c,T)}{\rho_d^0(T)} \,\exp(1-\phi_d (c, T) + \chi(1-\phi_d( c, T))^2), \\ \phi_d(c, T) &= (1-{c}) \frac{\rho( c, T)}{\rho_d^0( T)},
\end{align*}
with the diluent volume fraction $\phi_d$, the Flory-Huggins interaction parameter $\chi \in [0,1]$ and the gas constant $R = 8.3114$~J/(mol K). For the vapor pressure of the diluent $p_{vap}$ [Pa], we use for the acetone the Antoine equation 
\begin{align*}
\log \left(\frac{p_{vap}( T)}{p_0}\right) = A_0 - \frac{T_b}{{T} + T_c},
\end{align*}
with the parameters  $p_0 = 10^5$~Pa, $A_0=4.4245$, $T_b = 1312.253$~K, $T_c = -32.445$~K taken from \cite{ambrose:p:1974}. 
Because of the considered constant material properties (cf.\ Table~\ref{sec:ex_table:param}) the mixture density $\rho$ and the diluent volume fraction $\phi_d$ simplify to $\rho({c}) = \rho({c},{T})$ and $\phi_d( c)=\phi_d( c, T)$ . For simplicity we set $\chi = 0.5$.

\subsubsection{Thermal conductivity and diffusion coefficient}
For the thermal conductivity $C$ we choose the constant value $C = 0.2$ W/(mK). 

For the diffusion coefficient $D$ we use a model from \cite{verros:p:1999} that is based on the free volume diffusion model of \cite{vrentas:p:1977, vrentas:p:1977b}, i.e.,
\begin{align*}
D({c},{T}) =(1-\phi_d( c, T))^2(1-2\chi\phi_d(c,  T))D_0\exp\left(-\frac{V_1(1-{c}) + V_2{c}}{K_1(1-{c})(T_1+{T}) + K_2{c}(T_2+{T})}\right), 
\end{align*}
with the parameters
\begin{alignat*}{6}
V_1 &= 4.43\cdot10^{-4}, \qquad &V_2 &= 6.38\cdot10^{-4} ,\qquad &K_1 &= 1.86\cdot10^{-6}\text{ 1/K},\qquad
K_2 &= 5\cdot10^{-7}\text{ 1/K}, \\\ T_1 &= -53.33\text{ K}, \qquad &T_2 &= -240\text{ K}, & D_0 &= 3.6\cdot10^{-8}\text{ m$^2$/s}. &
\end{alignat*}

\subsubsection{Diluent and evaporation enthalpy}
Using the constant specific heat capacity for the diluent acetone $q_d^0$ (cf.\ Table~\ref{sec:ex_table:param}) we employ the caloric equation of state for the enthalpy with respect to the freezing point $T_f = 178.15$~K, see \cite{moore:b:1962},
\begin{align*}
h_d^0(T) = q_{d}^0(T-T_f).
\end{align*}
The evaporation enthalpy model $\delta$ is taken from \cite{majer:b:1985}, i.e.,
\begin{align*}
\delta(T) = \frac{E_0}{M_d}\exp\left(-\iota\frac{T}{T_\delta}\right)\left(1-\frac{T}{T_\delta}\right)^{\iota},
\end{align*}
with $E_0 = 46.95\cdot10^3$ J/mol, $\iota = 0.2826$ and $T_\delta = 508.2$ K.

\subsubsection{Dynamic viscosity}
For our setting the dynamic viscosity $\mu$ corresponds to the zero-shear viscosity $\eta_0$ [Pa s]. The model for $\eta_0$  is taken from \cite{gou:p:2003, gou:p:2004}, where an approach of \cite{krevelen:b:1990} is adopted, 
\begin{align*}
\eta_0(c,T) =
\begin{cases}
\eta_a P_n^{3.7} c^7\exp(\frac{\Delta E}{R T}), \quad & T \geq 1.2T_g(c,T),\\
\eta_0(c,1.2T_g(c,T))\cdot10^{-B_1(T-1.2T_g(c,t))(T_\eta+T-1.2T_g(c,T))^{-1}}, \quad & T_g \leq T < 1.2T_g(c,T),\\
\eta_0(c,T_g(c,T)), \quad & T < T_g(c,T).
\end{cases}
\end{align*}
Comparing the temperature $T$ to the glass transition temperature $T_g$ [K], the zero-shear viscosity follows an exponential decay for high temperatures, obeys the Williams-Landel-Ferry (WLF) model for temperatures near to $T_g$ and stays constant on the viscosity value at $T_g$ for low temperatures. The parameters are $\eta_a = 6.6\cdot10^{-13}$~Pa s, $T_\eta = 101.6$~K, $B_1 = 8.86$. The degree of polymerization of CA is $P_n = 200$ and the flow activation energy $\Delta E = 5.6940\cdot10^4$~J/mol. The glass transition temperature is prescribed according to the Kelley-Bueche equation
\begin{align*}
T_g(c,T) = \frac{B_2\phi_d(c,T)T_{g,d} + \phi_p(c)T_{g,p}}{B_2\phi_d(c,T) + \phi_p(c,T)},
\end{align*}
using $B_2=2.5$, the material glass transition temperatures $T_{g,p} = 468$~K, $T_{g,d} = 119$~K,  as well as the volume fractions $\phi_p$, $\phi_d$ of CA and acetone, respectively. Due to the considered constant polymer and diluent densities (Table~\ref{sec:ex_table:param}), the volume fractions simplify to $\phi_p(c) = c\rho(c)/\rho_p^0$ and $\phi_d(c) = (1-c)\rho(c)/\rho_d^0$. 

Due to the nonlinear form of the viscosity model including exponential growth, we cannot expect that the desired cross-sectionally averaged viscosity $\langle \mu(c,T) \rangle_{R^2}/\pi$ is well approximated by the viscosity of the averaged fields $\mu(\bar{c},\bar{T})$, cf.\ Remark~\ref{sec:model_remark:dyn_visc}. For reasons of accuracy we thus compute the profiles of the dynamic viscosity $\mu(c,T)$ in Step (2) of Algorithm~\ref{sec:numerics_algo1} and average them over the fiber cross-sections afterwards to obtain $\langle \mu(c,T) \rangle_{R^2}/\pi$. 
\begin{figure}[!htbp]
\begin{minipage}[c]{0.99\textwidth}
	\centering
	\includegraphics{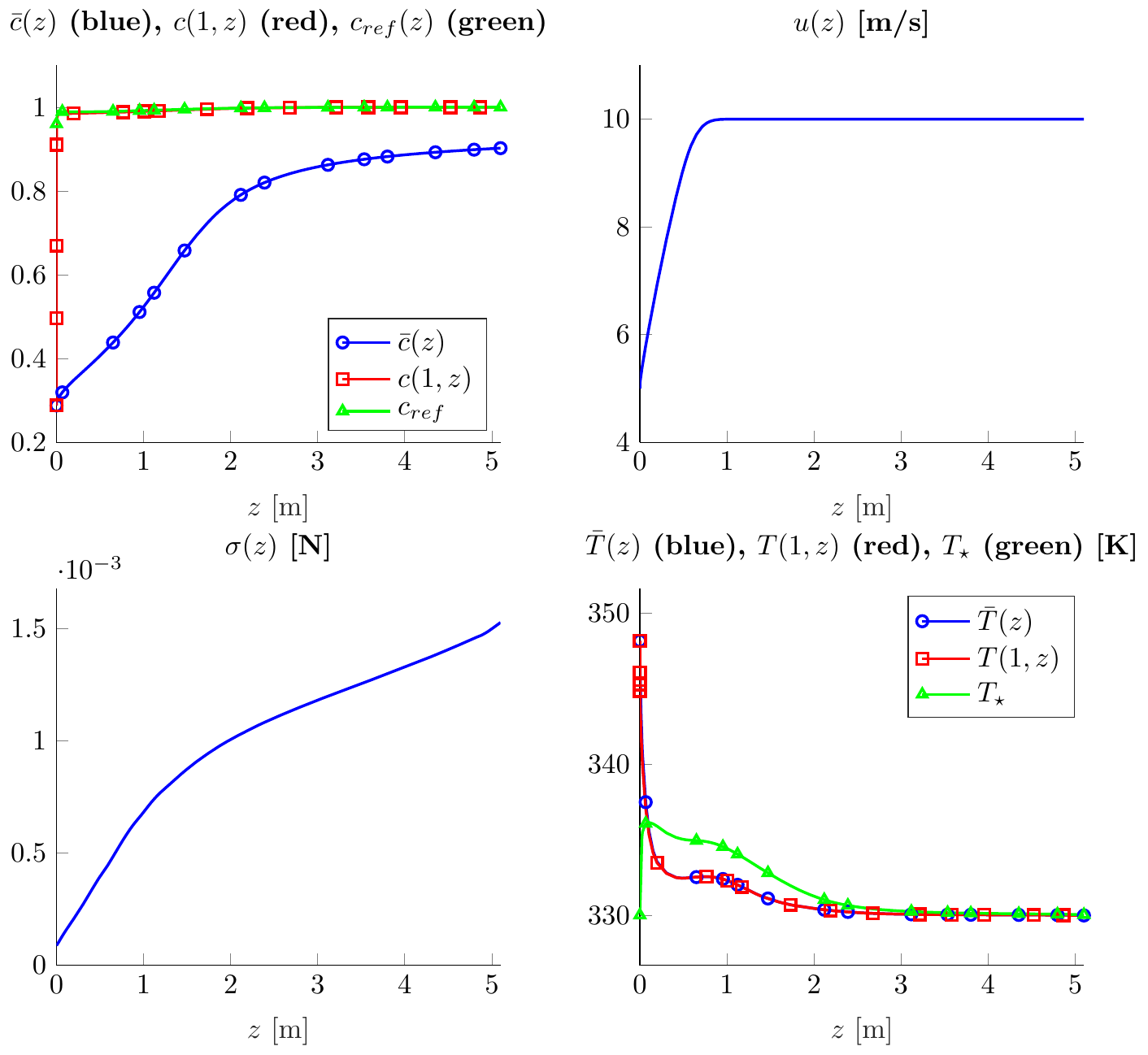}
\end{minipage}\vfill
\caption{Solution quantities of a centrally located fiber (cf.~Fig.~\ref{sec:ex_fig:spinneret}).  \textit{Top-left:} polymer (CA) mass fraction (averaged $\bar{c}$, at the fiber boundary $c\lvert_{r=1}$ and referential $c_{ref}$). \textit{Top-right:} scalar fiber speed $u$. \textit{Bottom-left:} tensile stress $\sigma$. \textit{Bottom-right:} temperature (averaged $\bar{T}$, at the fiber boundary $T\lvert_{r=1}$, air $T_\star$).}\label{sec:ex_fig:it30}
\end{figure}
\begin{figure}[!htbp]
\begin{minipage}[c]{0.99\textwidth}
	\centering
	\includegraphics{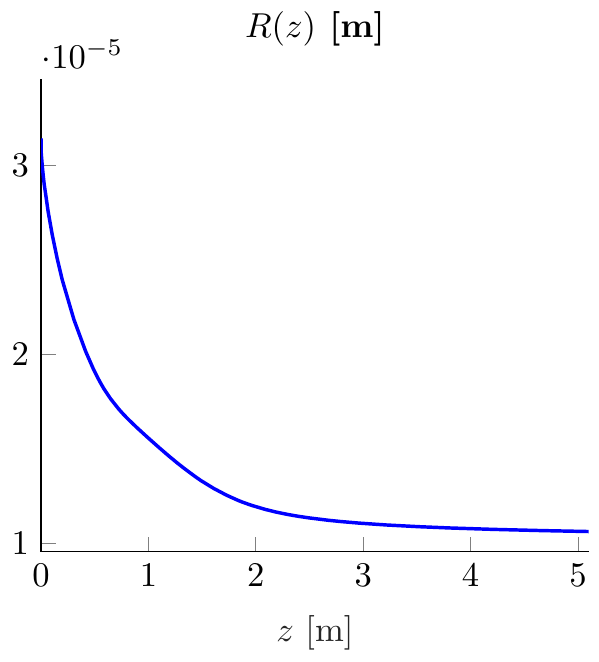}
\end{minipage}\vfill
\caption{Resulting fiber radius $R$ for the solution quantities in Fig~\ref{sec:ex_fig:it30}.}\label{sec:ex_fig:it30_radius}
\end{figure}
\subsection{Results} 
For the numerical treatment of the fiber dynamics we use the same mesh parameters and tolerances as in Sec.~\ref{sec:performance}. The solver for the airflow meets an accuracy of $\mathcal{O}(10^{-6})$. The break-up criterion of the fiber-air coupling algorithm satisfies an error tolerance $tol=10^{-5}$. For the considered industrial setup the fiber-air coupling algorithm converges after 30 iteration steps.

\begin{figure}[t]
\begin{minipage}[c]{0.49\textwidth}
	\centering
	\includegraphics{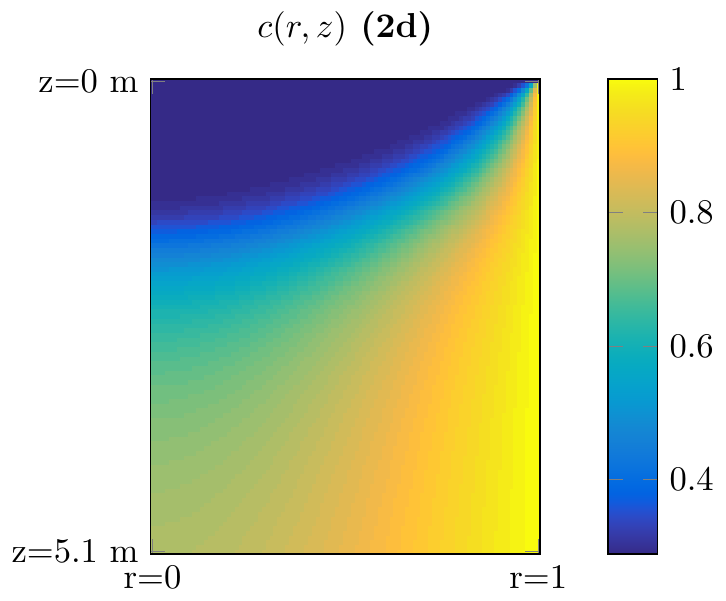}
\end{minipage}\hfill
\begin{minipage}[c]{0.49\textwidth}
	\centering
	\includegraphics{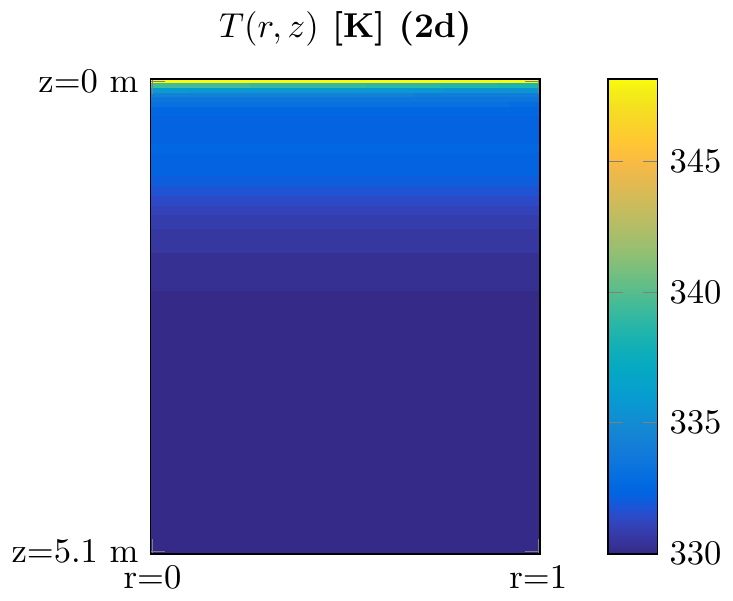}
\end{minipage}
\caption{Polymer mass fraction and temperature profiles of fiber solution (cf.~Figs.~\ref{sec:ex_fig:spinneret} and \ref{sec:ex_fig:it30}).}\label{sec:ex_fig:profiles}
\end{figure}

Since the solutions of the sixty fibers do not show visible differences, we illustrate -- as an example -- the solution behavior for the fiber that is centrally located in the spinning chamber (cf.\ Fig.~\ref{sec:ex_fig:spinneret}) in Fig.~\ref{sec:ex_fig:it30}. The cross-sectionally averaged CA mass fraction increases from $c_0 = 0.29$ at the inlet to $\bar{c} = 0.90$ at the outlet, indicating the evaporation of the acetone during the spinning process. The airflow directed contrary to the spinning direction of the fibers induces drag forces that act contrary to the fiber moving direction. Therefore, the spatial point where the fiber speed $u$ reaches the take up speed ($u_1=10$~m/s) is shifted from the center of the spinning device towards the nozzle, i.e., $u(z) = u_1$ for $z\geq 1$~m approximately. The tensile stress $\sigma$ increases correspondingly to the changes of $u$ and the viscosity $\langle \mu \rangle_\mathcal{A}$. Directly at the nozzle the fiber temperature drops down due to the immediate set in of the acetone mass transfer caused by evaporation, which is indicated by the rapid rise of the CA mass fraction at the fiber boundary. In a distance sufficiently far away from the nozzle the temperature approaches the temperature of the surrounding air $T_\star$. The fiber radius decreases from its initial value $R_0 = 3.14\cdot10^{-5}$~m to $R=1.06\cdot10^{-5}$~m at the fiber end (cf.\ Fig.~\ref{sec:ex_fig:it30_radius}). On the one hand, this fiber thinning is caused by the take up of the fiber at the bottom of the spinning device with a drawing speed that is higher than the initial fiber speed, i.e., $u_1>u_0$. On the other hand, the mass transfer due to acetone evaporation gives rise to further fiber thinning, in particular in the region of the spinning chamber where the fiber speed has already reached the take up speed ($z\geq 1$~m). Worth to investigate is the profile for the polymer mass fraction in Fig.~\ref{sec:ex_fig:profiles} that indicates an inhomogeneous CA-acetone distribution in radial direction. While the fiber surface is nearly diluent-free and thus completely dried out shortly away from the nozzle, the innermost part of the fiber contains much more diluent over its complete length and is not completely dried at the bottom of the chamber. In contrast there are no visible radial effects in the temperature (cf. Fig.~\ref{sec:ex_fig:profiles}).
\begin{figure}[t]
\includegraphics[height=7.5cm]{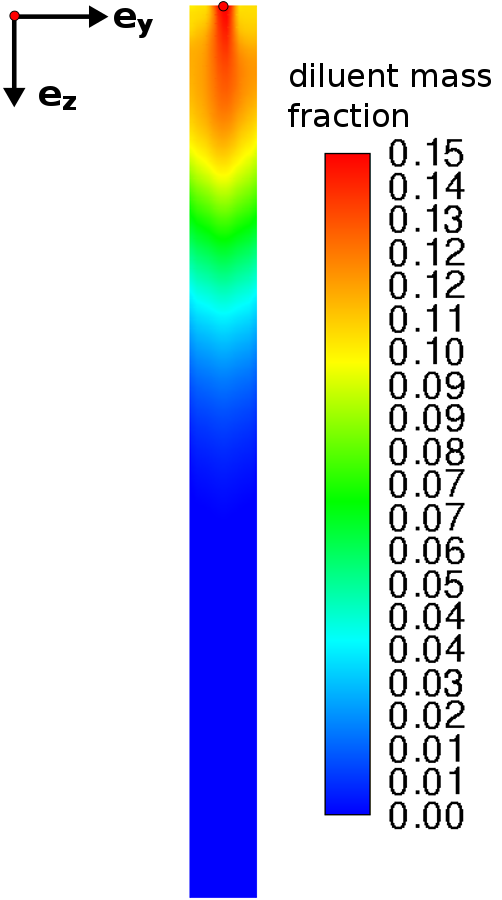}
\hspace*{1.5cm}
\includegraphics[height=7.5cm]{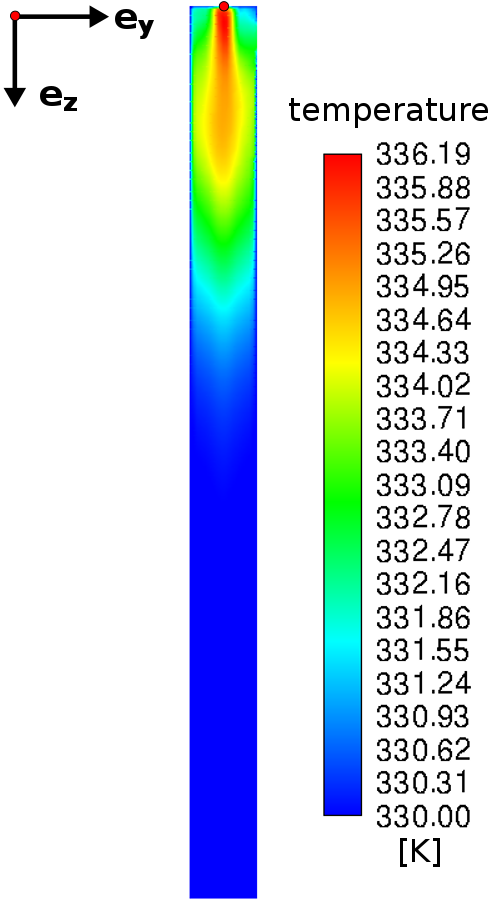}
\caption{Mass fraction of diluent acetone (\textit{left}) and temperature (\textit{right}) in the spinning chamber (after convergence of the fiber-air coupling algorithm).}\label{sec:ex_fig:geo}
\end{figure}
\begin{figure}[t]
\centering
\includegraphics{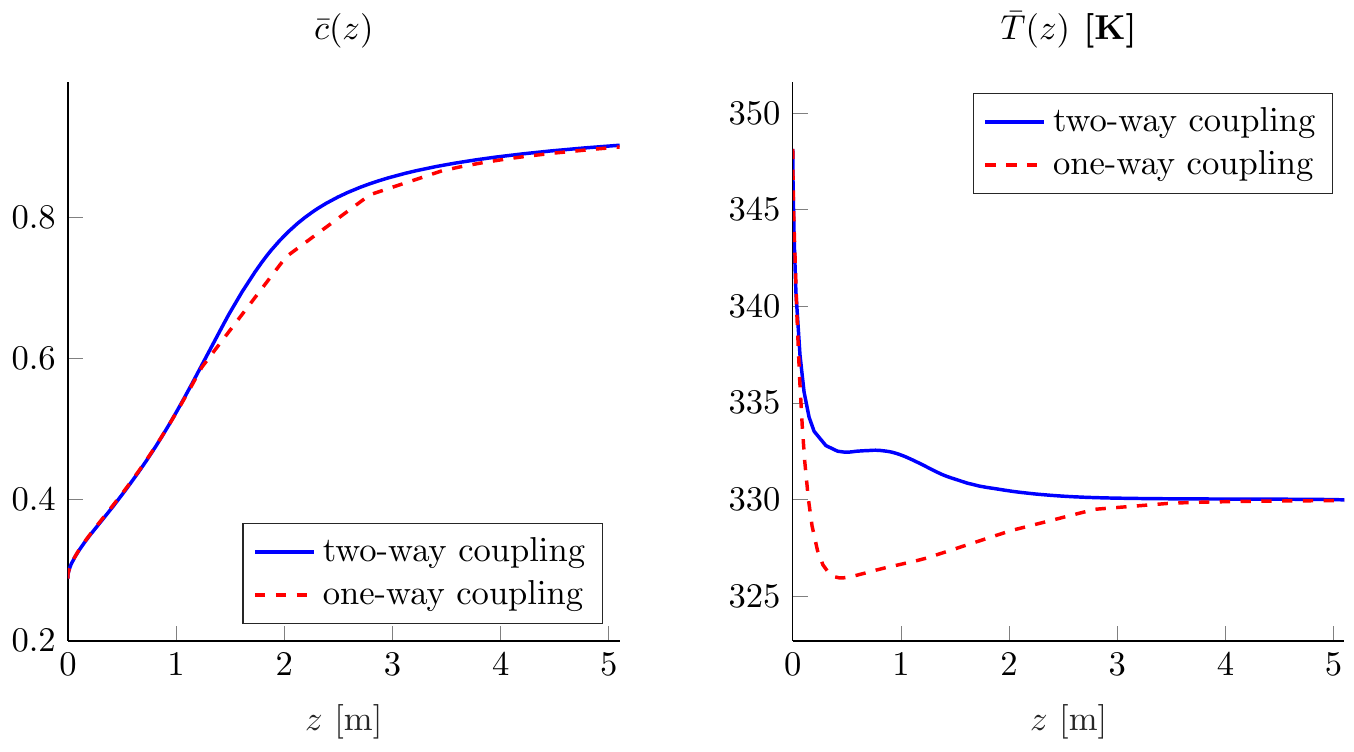}
\caption{Averaged CA mass fraction \textit{(left)} and temperature \textit{(right)} of a centrally located fiber after iteration 1 (one-way coupling) and after iteration 30 (two-way coupling) of the fiber-air coupling algorithm (cf.~Fig.~\ref{sec:ex_fig:spinneret}).}\label{sec:ex_fig:it1_30}
\end{figure}
\begin{figure}[t]
\centering
\includegraphics{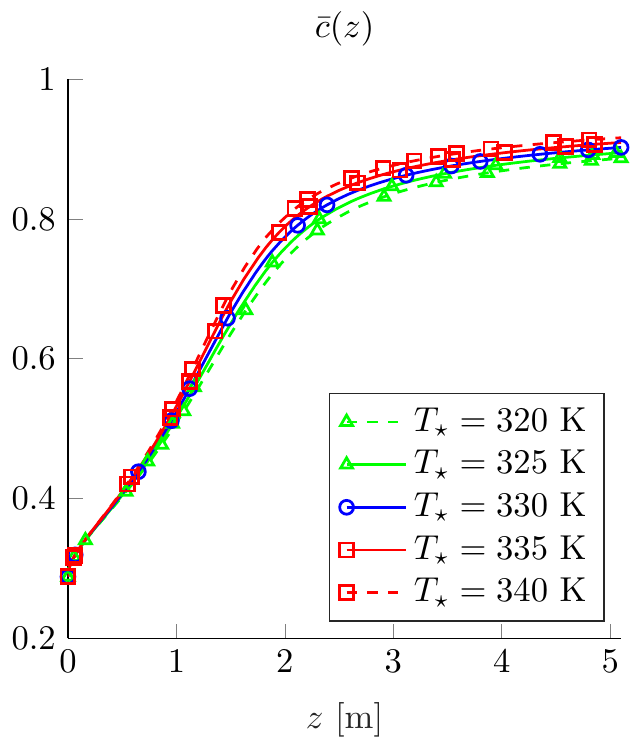}
\caption{Averaged CA mass fraction $\bar{c}$ for varying air inflow temperature $T_\star$.}\label{sec:ex_fig:it30_varTstar}
\end{figure}

The relevance of the effects due to the fiber-air interactions can be clearly seen in Fig.~\ref{sec:ex_fig:geo} and Fig.~\ref{sec:ex_fig:it1_30}. Due to the acetone evaporation the CA mass fraction in the fiber increases to the value $\bar{c} = 0.90$ at the end $z=L$ after the first iteration of the fiber-air coupling algorithm (one-way coupling). This means a ratio of acetone mass to CA mass of $0.11$ at the fiber end, which coincides with the value computed in \cite{sano:p:2001}. During the two-way coupling with the airflow the evaporated acetone increases its proportion in the surrounding air, such that the referential polymer mass fraction decreases from its initial value $c_{ref} = 1$ to its minimum $c_{ref} = 0.96$ directly at the nozzle ($z=0$ m), see Fig.~\ref{sec:ex_fig:it30}. Considering the temperature the influence of the fiber-air coupling becomes even more evident: in the region around the nozzle the fiber heats the surrounding air due to a positive relative temperature $T-T_\star > 0$. In turn this heated air affects the fiber temperature to decrease more slowly. The minimal fiber temperature after the first iteration is $\bar{T} = 325.94$~K, while after the 30th iteration it is $\bar{T} = 329.98$~K. Moreover, the air in the upper part of the chamber is heated up to $T_\star = 336.19$~K, which corresponds to a temperature increase of $6.19$~K compared to its initial value. This effect of increased air temperature in the nozzle region is also observed in \cite{sano:p:2001}, where the air temperature is not calculated but measured from experiments. Away from the nozzle evaporation effects become negligible and the temperature in the spinning chamber is not influenced significantly by the fiber temperature anymore. Vice versa the fiber is cooled down and reaches the air temperature in this region of the spinning chamber. Also here the temperature curve for our considered fiber after the first iteration (one-way coupling) is qualitatively in good agreement with the temperature curve given in \cite{sano:p:2001}, although our simulation shows a less drastic temperature drop at the nozzle. This difference might be due to varying rheological models.

The airflow in the considered dry spinning setup is characterized by the air inflow speed $\lVert \mathbf{v}_\star \rVert$ at the bottom of the device and its inflow temperature $T_\star$. Whereas the inflow speed has a negligible effect on the acetone evaporation, since the fiber speed $u$ is meanly determined by the take up speed $u_1$, the air temperature affects the evaporation of the solvent significantly, see Fig.~\ref{sec:ex_fig:it30_varTstar}. The higher the air temperature $T_\star$ is the more acetone evaporates out of the fiber and leads to a higher averaged CA mass fraction $\bar{c}$ remaining in the fiber, e.g. $T_\star = 320$ K yields $\bar{c} = 0.89$, $T_\star = 340$ K even $\bar{c} = 0.92$ at the fiber end.

Summing up, concerning one-way coupled fibers in air we achieve comparable results to \cite{sano:p:2001}. This justifies the transfer of our fiber model from academic axis-symmetric settings to real industrial setups. Moreover, our proposed procedure makes the simulation of industrial spinning processes with multiple fibers spun simultaneously and two-way coupled fiber-air interactions feasible. The computation time for the presented setup is around six hours.

\section{Conclusion}
The simulation of dry spinning processes requires the treatment of multiple fibers interacting with an surrounding airflow. Until now, there were no models and methods with the necessary efficiency to make such a simulation feasible. In this paper we deduced a dimensionally reduced viscous fiber model of one- and two-dimensional equations that covers the radial effects of mass fraction and temperature in combination with the tangential information of convective speed and tensile stress. Its good approximation qualities were shown in a comprehensive study regarding the solution of an underlying three-dimensional benchmark problem. The embedding of the two-dimensional equations in the solution framework of Green's functions and production integration method  led together with a robust continuation-collocation scheme for the one-dimensional equations to a very efficiently evaluable algorithm. The good performance of our proposed procedure allowed the simulation of an industrial dry spinning setup with two-way coupled fiber-air interactions in reasonable computation time. An extension of our model to curved fibers is straightforward following the concepts of \cite{panda:p:2008, marheineke:p:2009}. Consequently, our work builds a good basis for optimization and optimal design of dry spinning processes.


\end{document}